\documentclass[a4paper,fleqn,usenatbib]{mnras}

\usepackage[T1]{fontenc}
\usepackage{ae,aecompl}

\usepackage[dvipdfmx]{graphicx}	
\usepackage{amsmath}	
\usepackage{amssymb}	
\usepackage{multicol}
\usepackage{siunitx}
\usepackage{bmpsize}
\usepackage[anythingbreaks]{breakurl}
\newcommand{\mytilde}{\raise.17ex\hbox{$\scriptstyle\mathtt{\sim}$}}

\def\startdata{\if@table@not@headed\kill\caption{\\%
    \@tablecaption}\endhead\hline\endfoot%
  \fi%
}

\def\enddata{%
 \crcr 
 \noalign{\vskip .7ex}%
 \before@enddata 
 \endtabular 
 \setbox\pt@box\lastbox 
 \pt@width\wd\pt@box\box\pt@box 
}%

\newcommand{\aprx}{\raise.17ex\hbox{$\scriptstyle\sim$}}

\title[Transmission spectroscopy of WASP-52b]{A precise optical transmission spectrum of the inflated exoplanet WASP-52b}

\author[T. Louden et al.]{Tom Louden$^{1}$\thanks{E-mail: t.m.louden@warwick.ac.uk}, Peter J. Wheatley$^{1}$\thanks{E-mail: P.J.wheatley@warwick.ac.uk}, Patrick G. J. Irwin$^{2}$, James Kirk$^{1}$ and Ian Skillen$^{3}$\\
$^{1}$Department of Physics, University of Warwick, Coventry, CV4 7AL, UK\\
$^{2}$Clarendon Laboratory, Oxford University, Oxford, OX1 3PU, UK\\
$^{3}$Isaac Newton Group of Telescopes, Apartado de Correos 321, 38700 Santa Cruz de La Palma, Spain}

\date{Accepted XXX. Received YYY; in original form ZZZ}

\pubyear{2017}

\begin{document}
\label{firstpage}
\pagerange{\pageref{firstpage}--\pageref{lastpage}}
\maketitle
\begin{abstract}
We have measured a precise optical transmission spectrum for WASP-52b, a highly inflated hot Jupiter with an equilibrium temperature of 1300 K. Two transits of the planet were observed spectroscopically at low resolution with the auxiliary-port camera (ACAM) on the William Herschel Telescope (WHT), covering a wide range of 4000--8750\,\AA. We use a Gaussian process approach to model the correlated noise in the multi-wavelength light curves, resulting in a high precision relative transmission spectrum with errors on the order of a pressure scale height. We attempted to fit a variety of different representative model atmospheres to the transmission spectrum, but did not find a satisfactory match to the entire spectral range. For the majority of the covered wavelength range (4000--7750 \AA) the spectrum is flat, and can be explained by an optically thick and grey cloud layer at 0.1 mbar, but this is inconsistent with a slightly deeper transit at wavelengths $> 7750$ \AA. We were not able to find an obvious systematic source for this feature, so this opacity may be the result of an additional unknown absorber.

\end{abstract}




\section{Introduction}\label{sec:introduction}

Exoplanet science is rapidly developing into a mature discipline where detailed characterisation of individual planets is beginning to build a picture of the population of planets as a whole. The characterisation of exoplanet atmospheres, primarily through transmission and emission spectroscopy, has become a key topic. Hot Jupiter atmospheres have been shown to display a wide variety \citep{Sing2015}, but broadly can be categorised into atmospheres that are ``clear", displaying large sodium and potassium features (e.g. HD\,209458b), ``cloudy", which display muted, or no features at all at optical wavelengths, and ``hazy'' atmospheres, such as HD\,189733b where the dominant feature is a Rayleigh scattering slope caused by sub-micron scale scattering particles high in the atmosphere of the planet. While many high profile discoveries have been space based \citep[e.g.][]{Vidal-Madjar2003,Sing2008a,Kreidberg2014a}, increasingly, ground based telescopes have also been found to be suitable to the task for the detection and characterization of both narrow \citep[e.g.][]{Redfield08,Louden2015} and broadband features \citep[e.g.][]{Bento2013,Gibson2013,Gloria2015,Borsa2016}. While susceptible to their own systematic issues, progress has been made in identifying and improving the major problems, such as differential slit loss \citep{Sing2012} and upgrading the Linear Atmospheric Dispersion Corrector (LADC) on the FOcal Reducer and low dispersion Spectrograph (FORS2) \citep{Moehler2010,Sedaghati2015}.

Many exoplanets have been shown to present a flat, cloud dominated spectrum \citep[e.g.][]{Lendl2015,Gibson2013}. Data from \emph{HST} shows that broad wavelength coverage is required to characterize a hot Jupiter, as most appear flat over ranges of several hundred angstroms without the aid of high resolution measurements \citep{Sing2015}. So far, there is no obvious correlation between planetary parameters, e.g. equilibrium temperature, surface gravity, and host spectral type, though there are broad predictions about what kinds of condensates are available at different temperature ranges and their impact on observational properties \citep[e.g.][]{Wakeford2015,Parmentier2016}. The transmission spectrum of a planet is an emergent property of many branches of atmospheric physics, that likely depends quite sensitively on the specifics of heat transport and vertical mixing in each individual planet. It will probably be necessary to analyse many more systems before the underlying trends begin to emerge.

Ground based observations have the advantage of being relatively cheap to perform in large number, compared to space based observations, and they also fill an important role in validating the results from \emph{HST} and \emph{Spitzer}, where exoplanet observations are known to suffer from the idiosyncratic systematic errors of these instruments \citep[e.g.][]{Evans2015,Wakeford2016}. The way these systematics are treated can have a large impact on the resulting transmission spectra, and many early results are controversial \citep[e.g., the detection of a thermal inversion on HD\,209458b][]{Knutson2008a,Line2013}. Independent detections  with a wider range of instruments would remove some of the uncertainty stemming from these systematics concerns. 

Additionally, wavelength dependent trends in the data injected by, for example, stellar activity could be mistaken for real atmospheric features on the planet. The repeatability of a purported signal over several observations is a powerful diagnostic, giving immense value to having multiple nights on target. Ground-based, broad-band transmission spectroscopy therefore remains a vital tool in building an accurate picture of exoplanet atmospheres.

WASP-52b is a Saturn-mass planet with an inflated radius that orbits a K type star on a 1.75 day orbit \citep{Hebrard2012}. WASP-52b is an excellent target for transmission spectroscopy due to its large scale height, relatively deep transit, and the presence of a nearby bright comparison star. The low density of WASP-52b gives it a scale height of 700 km, which is over three times greater than HD\,189733b, despite having a similar equilibrium temperature of 1300 K. Although it orbits a comparatively faint star ($V_{mag}=12$), this high scale height means that the Signal-to-Noise achievable on atmospheric features is comparable with HD\,189733b, assuming that white noise is the dominant noise component, and potentially better where red noise is a limiting factor. The similarity of the planet's equilibrium temperature and parent star spectral type to HD\,189733b make WASP-52b an interesting test of the variations in exoplanet atmospheres.

We present two high precision spectroscopic transits of WASP-52b. This paper is organised as follows: a summary of the observations are given in Section \ref{sec:Observations}, the reduction of the data is described in Section \ref{sec:extraction}, and a short description of the Common mode Gaussian process detrending analysis is presented in Section \ref{sec:gps}. The system parameters are refined in Section \ref{sec:system} and a low resolution spectrum is presented with a discussion of its features in Section \ref{sec:Discussion}.

\section{Observations}\label{sec:Observations}

WASP-52 was observed for 2 full nights on August 22 and 29th 2014 using the auxiliary-port camera (ACAM) on the William Herschel Telescope (WHT). ACAM was used in single slit spectroscopy mode with a custom made extra wide slit of 27". This was to minimise slit losses, which have been found to be a limiting factor in past observations with other instruments \citep{Sing2012}, though a slit is still of course necessary to avoid contamination from other stars in the field of view and to lower sky background contribution. A slit of 27" arcseconds was chosen after an earlier analysis of the Point Spread Function (PSF) of the instrument in an earlier dataset, with the goal of there being less than 0.1\% slit losses.
A comparison star with $V_{mag}=13$ and a similar spectral type located 3 arcminutes away, well within the unvignetted slit length, was selected to perform differential spectroscopy. The slit was rotated such that both stars were at the center.
An exposure time of 100 seconds was used to keep the peak flux well below saturation at all times. The windowed observing mode reduced the readout time to 11 seconds, giving a duty cycle efficiency of 90\%. In order to keep the stars centred in the slit, the data were reduced in realtime and the resulting spectra were cross-correlated to produce guiding offsets, using the position of the spectral traces in the X-direction on the chip, and the positions of strong absorption features in the spectra in the Y-direction. Pointing corrections were entered manually at intervals of approximately 20 minutes throughout the observations, with typical amplitudes of 0.1 pixels, these small and frequent corrections prevented the formation of a sawtooth pattern in the trace positions. A small defocus was applied to the telescope (\aprx 2") to spread the light over a larger number of pixels to reduce the contribution of flat fielding errors, and to create a more stable PSF that is less effected by variable seeing.

\section{Method}\label{sec:method}

\subsection{Extraction}\label{sec:extraction}

\begin{figure}
\begin{center}
\includegraphics[width=\columnwidth]{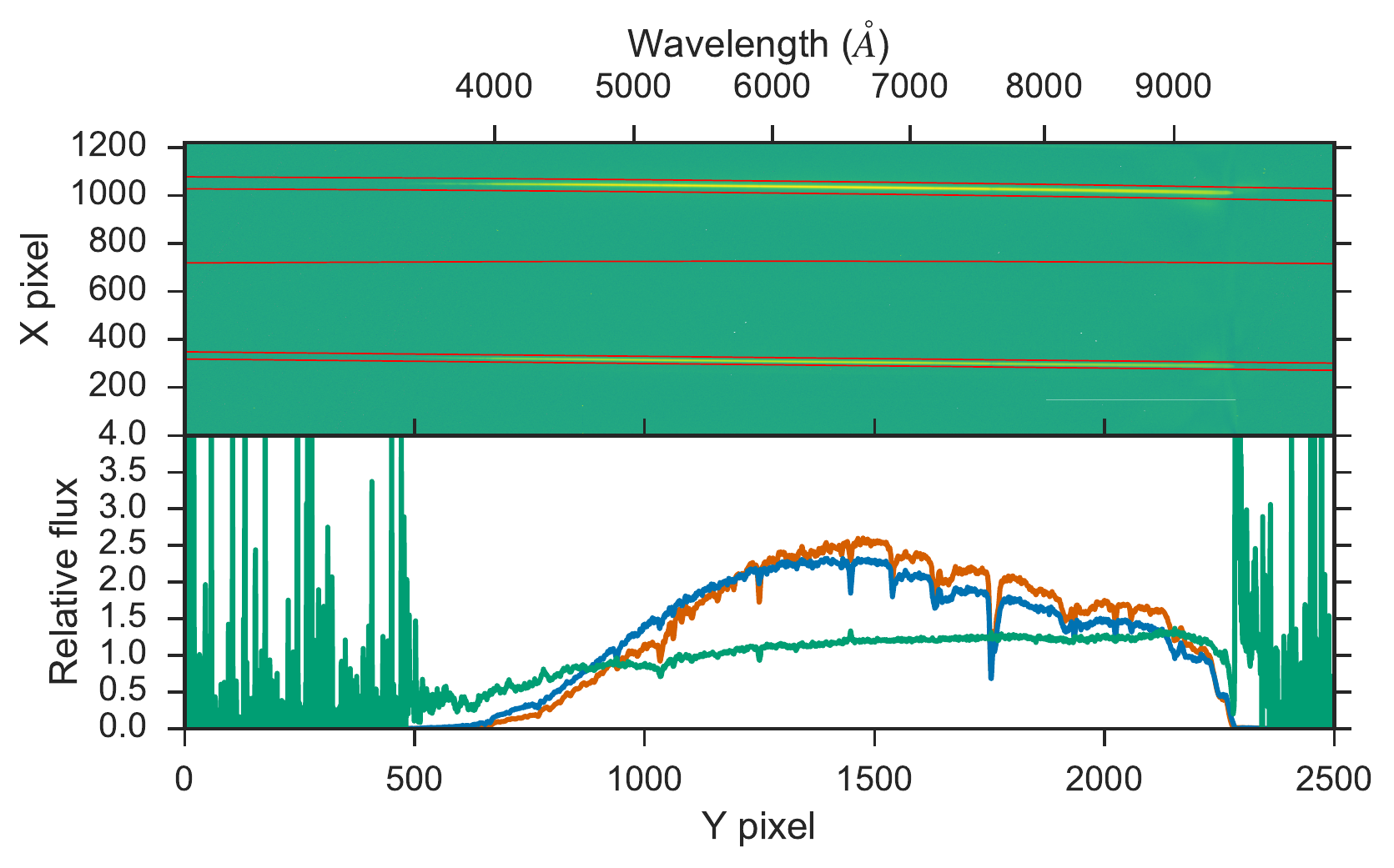}
\caption{Spectral trace for target and comparison star. Top: The spectrum image after background subtraction. For display purposes, the image has been normalised by dividing each pixel row by its standard deviation, such that the colour scale represents the significance above the background. This is only for visual clarity, and is not a part of the actual data reduction. The extraction region for each trace is represented by red lines. The central line indicates the distance between the two traces, as a function of pixel position. Bottom: Normalised extracted spectra for target (orange) and comparison (blue). The ratio between the two is plotted in green. The usable region of the spectrum between pixels \aprx 600 and 2200 corresponds to wavelengths of \aprx 4000--9000 \AA, the wavelength solution is marked on the top panel.}
\label{fig:extract_region}
\end{center}
\end{figure}

The reduction was carried out with custom scripts written in Python. Standard bias corrections were made. However, it was not possible to flat field the data satisfactorily with the sky flats taken due to differential vignetting between the imaging and spectral modes of ACAM. Spectroscopic dome flats did not contain enough blue light to be useful across the whole spectral range. Due to the large number of pixels used during the analysis, and the small spatial scale of flat field variations compared to the variations within the spectra the flat fielding errors are not expected to be an important error source, so the reduction was carried out without flat fielding.

The combination of vignetting towards the edge of the chip and the broad wings of the brighter comparison star made it difficult to judge where to place sky background traces. However, the background was found to vary smoothly and predictably over the rest of the chip, so we decided to fit this function directly using a third order iterative sigma clipped polynomial and subtract it from the image on a row by row basis (here rows are perpendicular to the dispersion direction). This approach has the advantage of leveraging information from a larger number of pixels, which increases the accuracy of the background estimate. There is a small curvature in the sky lines in the x-direction due to the alignment of the instrument, but it is significantly less than the width of the lines, so does not effect the background subtraction.

We fit the spectral trace for the target and reference star with an iterative cubic spline. Several extraction widths were tested, and it was found that a width of 20 pixels provided the lowest scatter in the final lightcurves on both nights. The adjustments to the guiding kept the brightest region of the spectrum dispersed over the same set of pixels, but the spectral trace was found to rotate slowly throughout the night due to differential refraction, meaning the extreme blue end of the spectrum traversed a slightly larger distance throughout the observation and may suffer an increased noise budget due to the lack of flat-fielding. The bluest end of the usable spectrum at 4000 \AA (Pixel row 674) was found to drift by two pixels throughout the night, while the central regions and the red end drift by less than a pixel (0.25").

An example spectral extraction is shown in Figure \ref{fig:extract_region}. The colour scale has been selected to emphasise the target and reference star trails. It is clear that the resulting spectra and their ratios are well behaved with high S/N between pixels \aprx 600--2200, which are found to correspond to wavelengths \aprx 4000--9000 \AA.

Persistent artefacts are visible in the images, which we suspect are caused by fringing. These become visible beyond \aprx 8750 \AA, and make it more difficult to trust the spectra extracted in this region. Bluewards of 4000 \AA\ a combination of low intrinsic flux, detector response, and instrument vignetting drop the count rate below the background, making a total usable range of 4000--8750 \AA.

An accurate and stable wavelength calibration is important, as errors in wavelength calibration can cause systematic bluewards and redwards slopes in the transmission spectrum that can be mistaken for spectral features in the atmosphere of the planet. Due to the very wide slit, arc calibration frames taken with the same slit produce very few useful features, particularly at the blue ends of the spectrum where the CuAr and CuNe calibration lamps available at the WHT produce very few lines. Using a smaller slit and removing the offset in slit positions is possible, but we found that the wavelength solution drifted noticeably over the 8 hour observation period, which is caused by flexure of the ACAM instrument. Breaking science observations and changing the instrument setup to take regular arcs to recalibrate is undesirable, so we elected instead to self-calibrate the science spectra using telluric and strong stellar features. At the resolution of these observations, the velocity difference between the stellar and telluric features is unimportant. The line positions are measured by fitting a Gaussian profile, which well describes the instrumentally broadened profiles. The central pixel co-ordinates of these features are used to generate a wavelength solution using a 3rd order polynomial fit. Each spectrum is then re-sampled onto a common wavelength grid using pysynphot\footnote{Distributed as part of the Space Telescope Science Institute
stsci\_python package, \\ www.stsci.edu/institute/software\_hardware/pyraf/stsci\_python}, which conserves flux. This produces a stable and consistent wavelength solution for the entire set of observations.

Obvious cosmic ray hits are removed from the individual spectra at this stage. A sliding Gaussian weighted average with the FWHM of the instrument is run along the spectra, and points that are $5\sigma$ or greater discrepant with this value are replaced by it. Typically only a small number of points per spectrum are affected.

To generate a light curve we took the sum of the flux in the desired wavelength range at each time point for both target and comparison, and then took the ratio. To reduce computing time in the systematic noise model analysis (described in section \ref{sec:gps}) we performed a basic fit first to remove extreme outliers. We fit a simple transit model \citep{Mandel2002} with a second order polynomial baseline fit, and removed any points greater than $5\sigma$~from the model. Most lightcurves did not require more than one point to be removed. Each lightcurve was then normalised such that the median of the out-of-transit data is 1.

\begin{figure}
\begin{center}
\includegraphics[width=\columnwidth]{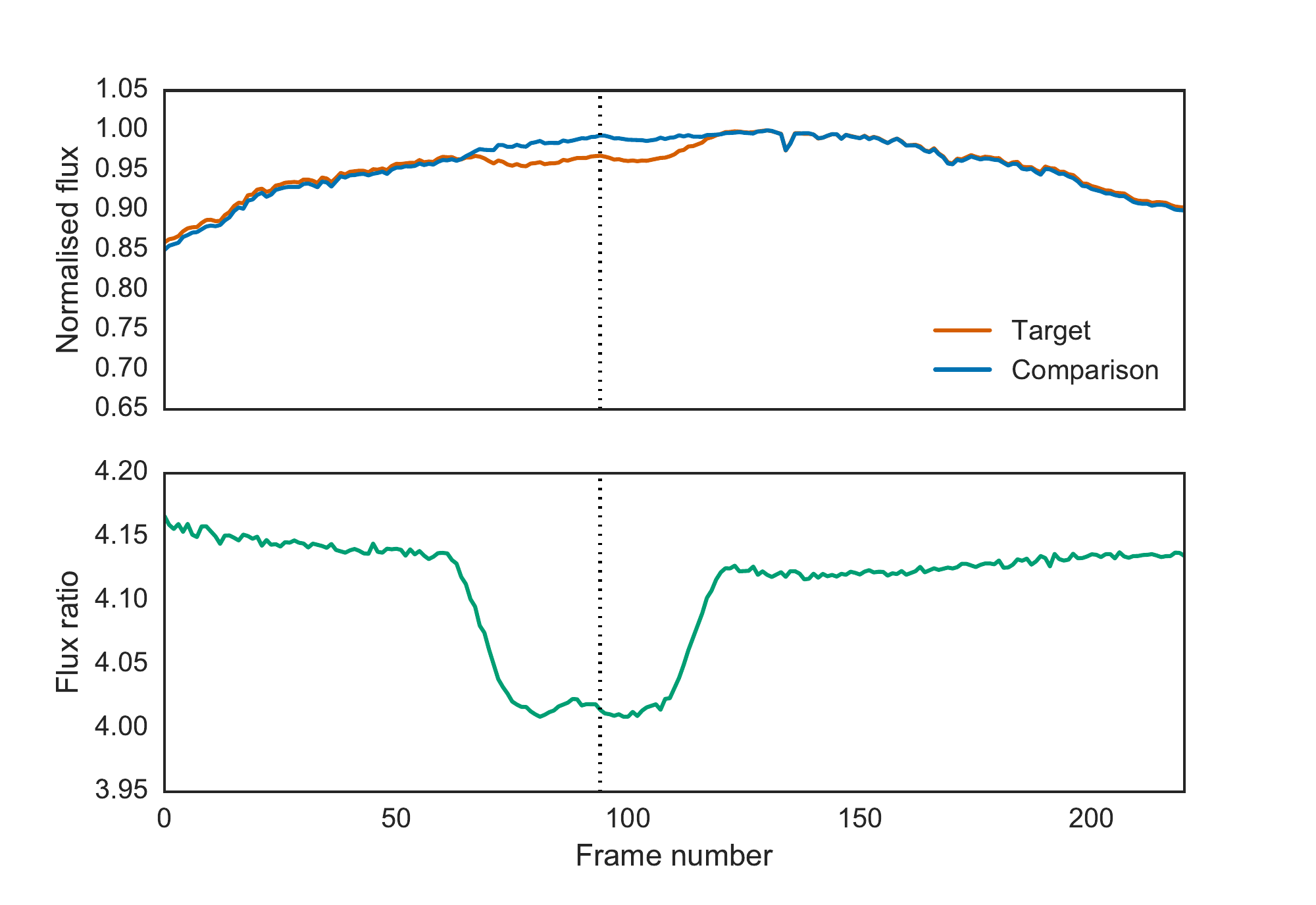}
\includegraphics[width=\columnwidth]{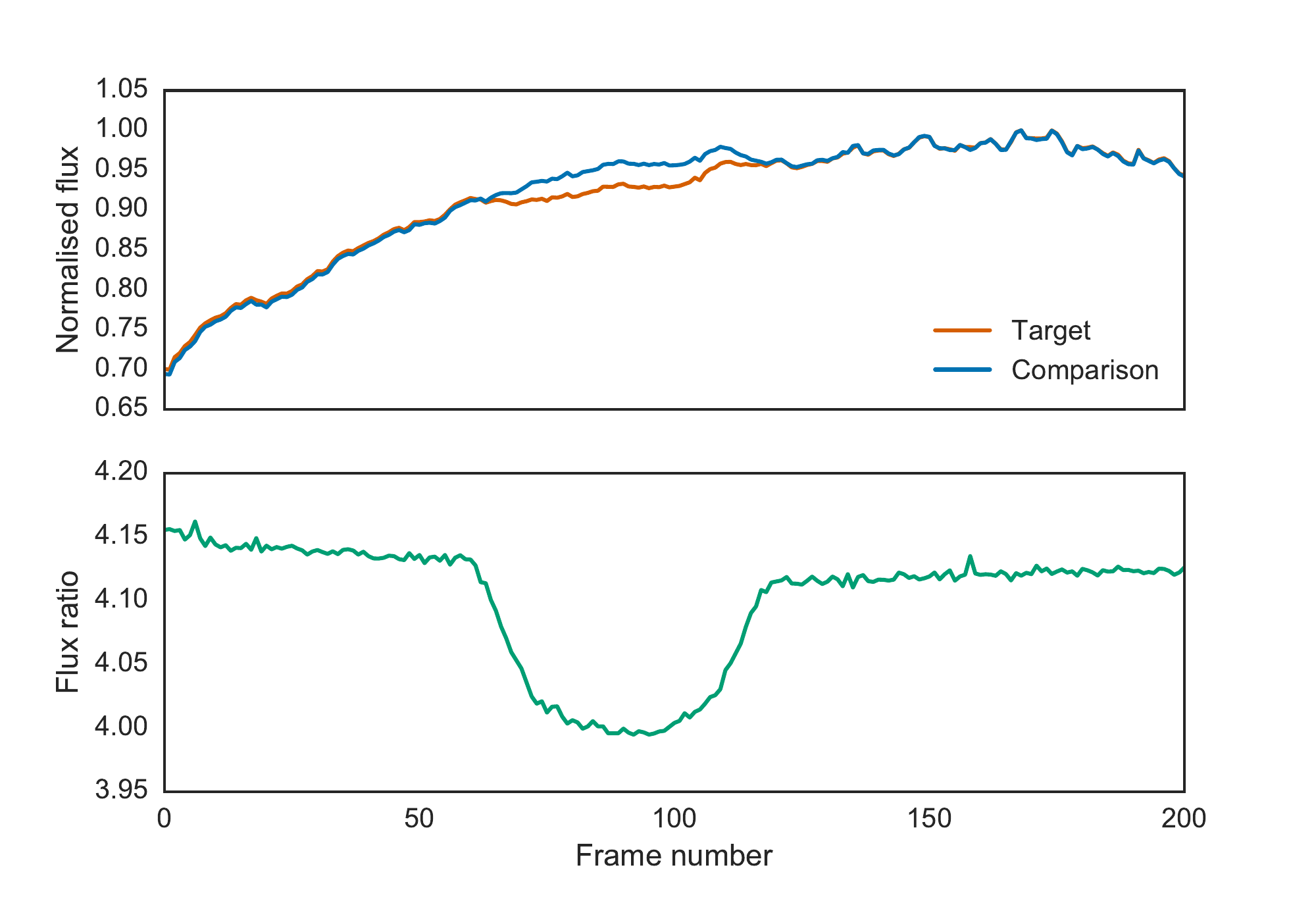}
\caption[caption]{Raw white lightcurves for the flux from WASP-52 (orange) and the comparison star (blue), and their ratio (green), for night 1 (top) and night 2 (bottom). For night 1 a dashed line indicates the position of a feature visible in both of the raw lightcurves which is removed by the ratio, in order to distinguish it from the nearby feature in the light curve ratio, which is most likely astrophysical in nature.}
\label{fig:rawlightcurves}
\end{center}
\end{figure}

The raw white-light (4000-8750 \AA) lightcurves for the two nights and their ratio are shown in Figure \ref{fig:rawlightcurves}. On both nights the dominant trend in the extracted data is the airmass term, which is mostly removed by taking the ratio. Small fluctuations in transparency are likewise shared between target and comparison, so are not present in the ratio.

During the first night there is a feature close to the middle of transit, this feature is discussed in more detail in section \ref{sec:spotcross}. This potential spot-crossing feature is distinct from a bump in the data that is closer to the center of the transit, which is present in both the target and comparison lightcurves and is removed by taking the ratio. The location of this non-astrophysical bump is marked in Figure \ref{fig:rawlightcurves}.

\begin{figure}
\begin{center}
\includegraphics[width=\columnwidth]{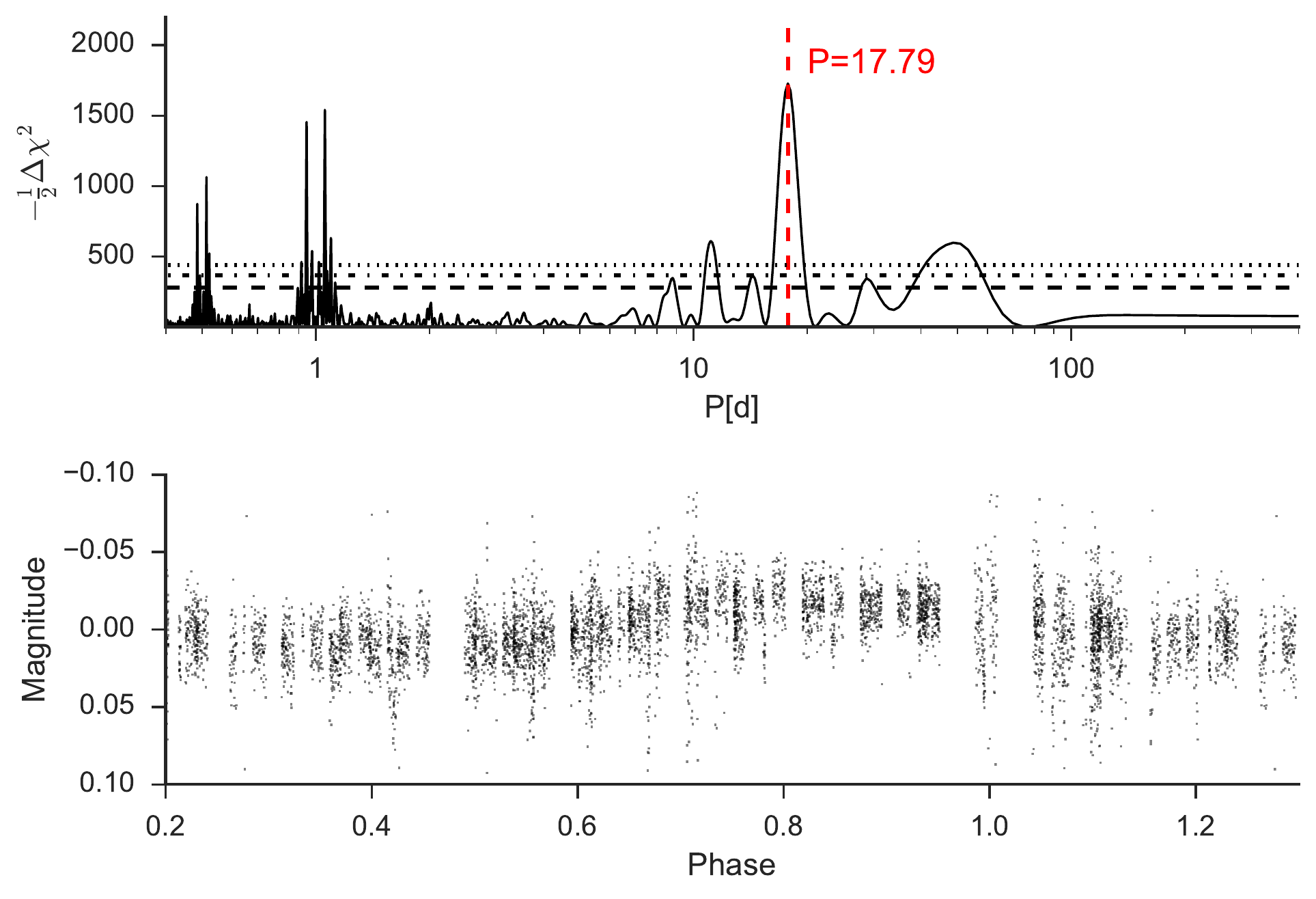}
\caption[WASP-52 periodogram of stellar activity]{Top: A periodogram for the most recent season of WASP data on WASP-52. The stellar rotation period is clearly detected at 17.79 days. The one, two and three $\sigma$ FAP limits are indicated. Bottom: Photometric data from WASP folded on the detected period of 17.79 days.}
\label{fig:periodogram}
\end{center}
\end{figure}

\subsection{Fitting procedure}\label{sec:fitting}

In order to generate the final transmission spectrum for each night we follow several steps, each of which will be expanded on in detail. First, the transit and systematics model (described in \ref{sec:gps}) is fit to the whole of the usable spectral range (4000--8750 \AA). We then divide the spectrum into 250 \AA\ spectral bins and fit the transit model to each separately, after removing the best-fit systematics model for that night, and fixing the system parameters to the values found for the white light curve. This removes the dominant systematics of the whole spectral range from the individual bins, which significantly lowers the error bars.

A Markov Chain Monte Carlo approach is used to maximise model parameter likelihood and estimate their errors. To perform the MCMC integration, we use the \textsc{emcee} code of \citet{Foreman-Mackey2013}\footnote{available at https://github.com/dfm/emcee}, which implements the affine-invariant ensemble sampler described in \citet{Goodman2010}.

To fit the transit light curve, we use the analytic models described in \citet{Mandel2002}. The eccentricity is fixed to 0, and at each step we draw the period from a distribution with mean and standard deviation values from \citet{Hebrard2012}. The inclination ($i$), system scale ($a/R_*$), radius ratio ($R_p/R_*$) and quadratic limb darkening co-efficients ($u1$,$u2$) are free parameters, as well as the parameters of the systematics model. LDTK (Limb Darkening Toolkit) \citep{Parviainen2015} was used to calculate the priors on the limb darkening parameters for each wavelength range. 

Each fit has 50,000 steps with 100 walkers, with the first 10,000 steps discarded as burn in. Convergence was checked by visual inspection of the walker chains and the evolution of their mean and standard deviation for each parameter.

The light curves for the full spectral range for the two nights are first fit simultaneously in order to calculate the best fitting system parameters, but at all other times the two nights are treated independently.

\subsection{Gaussian process and white noise model}\label{sec:gps}

Gaussian processes are a generalized class of functions that can be used to model correlated noise in time series data. They have a wide range of applications, but have increasingly become popular for accurate parameter estimation for exoplanet transits \citep[e.g.][]{Gibson2013,Parviainen2015a,Evans2015}. The covariance between data points is modelled with a ``kernel", which has a simplified functional form. Including a Gaussian process (GP) noise model allows one to effectively marginalise over the sets of functions that can represent the systematic noise in the data that do not hold any scientific interest. In our case, we use the Gaussian process to model everything in the data that is not explained by the transit model.

In this work the Matern 3/2 kernel is used, following \citet{Gibson2013a} who found it performed marginally better than other common kernels for transmission spectroscopy. The Matern kernel is similar to the commonly used squared exponential function, but is more sharply peaked, resulting in more flexibility at short timescales:
\begin{equation} \label{eq:kernel}
k(t_i,t_j) = \alpha \left(1 + \sqrt{3}\left(\frac{t_i - t_j}{\tau}\right)\right) \exp\left(-\sqrt{3} \left(\frac{t_i - t_j}{\tau}\right)\right)
\end{equation}
where $\alpha$ is the amplitude of the kernel, and $\tau$ represents a characteristic time-scale of the correlation. In practice we fit for the logarithm of these values, as they are expected to vary over several orders of magnitude.

The python package \textsc{george}\footnote{availible at https://github.com/dfm/george} \citep{Ambikasaran2016} is used to efficiently calculate the Gaussian process covariance matrices and the Likelihood at each Markov chain step.

There was a slightly higher level of white noise in the data than could be explained by photon noise alone, presumably due to scintillation. To account for this, and other sources of white noise, a Gaussian Process model will often include an additional term, $\sigma_w$, to increase the calculated error bars by a fixed amount.

For the spectrally resolved fits we found that in the bluest regions the level of additional white noise required was time varying, so we chose to take a similar approach to \citet{Gibson2013} and model the white noise as a smoothly time varying function. This makes sense, since an enhancement to the white noise, most likely due to changing atmospheric conditions, is not necessarily constant. The white noise term in equation~\ref{eq:kmatrix} is modelled as
\begin{equation} \label{eq:wnoise}
\sigma(t_i) = \sigma_a \exp\left(-\frac{t_i}{\sigma_b}\right) + \sigma_c
\end{equation}
where $\sigma_a$, $\sigma_b$ and $\sigma_c$ are additional terms in the model fit.

For consistency, we fit all the spectral lightcurves with this white noise model. We did not think it was necessary to re-fit the white light curves with this new model, as their residuals do not vary significantly with time. We find that for most wavelength regions the posteriors using the time varying white noise model are no different to those with the fixed increase, but in the bluest channels, where scintillation is expected to have the greatest effect, the more accurate characterisation of the noise properties slightly narrows the credible intervals of the transit depth, but only marginally changes the mean values.

\subsection{Application of the model}

Due to the presence of the spot-crossing event on the first night, we chose to initially fit the two nights separately to ensure that the Gaussian process systematic model was capable of fitting out the event without introducing bias into the other parameter values. Comparison of the parameter values satisfied us that, apart from a small difference in absolute depth which is expected from changes in activity levels, the parameters are consistent on both nights (see Table \ref{tab:system}). The final parameter values are calculated by fitting the two nights simultaneously.

Ultimately, what we are interested in is the \emph{relative} depth of the wavelength channels, or their conditional probability distribution with respect to each other in order to build a transmission spectrum. Systematic noise sources, such as guiding instability or changes in atmospheric transparency will be the same, or subject to a small scaling factors between bands. It is therefore incorrect to treat a resulting transmission spectrum as a set of independent data-points, and their calculated error bars will be higher than the intrinsic scatter between them. This could also lead to unrealistic model fits, since most goodness of fit statistics, such as $\chi^2$, assume that the datapoints are independent.

We use the white light curves for each night to remove the common noise trends. A transit model with our Gaussian process noise model is fit to the normalised white light curve for each night. The best Gaussian process model is then taken and assumed to represent the systematic noise and trends in the data and the spectroscopic lightcurves are divided by this model, thereby removing the systematic components that are common.

\renewcommand{\arraystretch}{1.5}
\begin{center}
\begin{table*}{
\caption{Derived system parameters for WASP-52b. The combined fit includes constraints from \citet{Kirk2016}and \citet{Hebrard2012}.}
\begin{center}
\begin{tabular}{l c c c c c}
\hline
\hline
Parameter & Symbol & Night 1 & Night 2 & Combined & Unit\\
\hline
Epoch (JD$_{UTC}-2450000$) & $T_0$ &$6892.53861\pm 0.00029$&$6899.53749\pm 0.00016$& $5793.676184 \pm 0.000033$ & days\\
Epoch (BJD$_{TBD}-2450000$) & $T_0$ &$6892.54464 \pm 0.00029$&$6899.54374 \pm 0.00016$& $5793.682045 \pm 0.000033$ & days\\
Orbital period & $P$ & - & - & $1.74978089 \pm 0.00000013$ & days\\
Inclination & $i$ &$85.33^{+0.22}_{-0.23}$& $85.28^{+0.18}_{-0.17}$ &$85.32^{+0.14}_{-0.14}$& degrees\\
System Scale & $a/R_{*}$ &$7.23^{+0.12}_{-0.13}$&$7.176^{+0.089}_{-0.087}$&$7.200^{+0.076}_{-0.071}$&-\\
Radius Ratio & $R_{P}/R_{*}$ &$0.1741^{+0.0063}_{-0.0054}$&$0.1639^{+0.0030}_{-0.0030}$&-&-\\
\end{tabular}
\end{center}
\label{tab:system}
}
\end{table*}
\end{center}
\renewcommand{\arraystretch}{1.0}

\subsection{Spot crossing event}\label{sec:spotcross}

Due to the changing activity levels of the star, it would not a priori be expected that the depth, and hence inferred radius should be consistent between the two nights, as the stellar surface brightness distribution may be inhomogenous due to spots or plages. Indeed, we find that on the night with the apparent spot crossing event the implied transit depth is slightly deeper than the other, although the difference in absolute depths between the two nights is not formally significant, at $0.0102 \pm 0.0067$. Taking these results at face value, they would imply that the stellar surface was on average 0.4\% dimmer on that night. Whilst we unfortunately did not have access to contemporaneous monitoring of WASP-52 during the observation period, \citet{Hebrard2012} found that during their second observing season WASP-52 displayed a $16.4 \pm 0.04$ day period with an amplitude of 0.89\%, though there was no significant periodicity found in the first season. The separation between the two transits we observed is 7 days, or roughly half of the rotation period of the star -- so it might be expected that the observed spot coverage fraction could be different, despite the observations being close in time.

Since \citet{Hebrard2012} carried out their analysis, an additional season of SuperWASP-North \citep{Faedi2011} data on WASP-52 has become available. We carried out a harmonic analysis to assess how the activity levels vary between seasons. We used a generalized weighted Lomb Scargle periodogram \citep{Zechmeister2009} to search for significant periods in the data, which were taken between HJD 2455400 -- 2455800. We generate False Alarm Probability (FAP) power levels by using the night shuffling technique presented in \citet{CollierCameron2009} and \citet{Maxted2011}. Randomly sampling the nights instead of individual datapoints preserves correlated noise in the data, whilst removing any periodicities greater than 1 day, this gives a more realistic and conservative estimate of the FAP. The results are plotted in Figure \ref{fig:periodogram}, there is a highly significant peak at 17.79 days in the most recent season. To estimate the errors on the period and the amplitude we fit a sine curve to the data and ran an MCMC of 10,000 steps, centred around the peak of the periodogram, with a burn in of 1000 steps. We find that the fractional amplitude is $0.0142^{+0.0003}_{-0.0003}$ and the period is $17.79^{+0.05}_{-0.05}$ days. The amplitude is significantly higher than during the first and second seasons observed with WASP, which may be indicative of activity cycles. The period is 1.4 days longer than the period found in the second season, which is likely caused by the active regions occurring at different latitudes and differential rotation.

Despite the lack of contemporaneous coverage, the difference in absolute depth between the two nights is therefore consistent with the historical activity levels of the star. Note that the depth on the uncontaminated night is consistent with both the original published value, and the faculae model in \citet{Kirk2016}, but is inconsistent with their spot model. All other parameter values agree to within one sigma, and there is no sign of correlation between the GP hyper-parameters and transit parameters (except depth) in the MCMC corner plots 
The significant difference between the spot model in \citet{Kirk2016} and both the discovery paper and this work strongly supports their conclusion that their data shows the planet crossing faculae rather than spots.

Since the spot crossing event appears in our data to be grey to within the level of precision, the approach of removing the common Gaussian process noise model is preferred over attempting to model the spot directly. This would add additional model parameters and degeneracies, and would not improve the level of precision attained in the recovered transmission spectrum. Inspection of the GP model for night 1 (Figure \ref{fig:whitelcs}) reveals at least 2 distinct components which indicate that the idealisation of a single ``spot crossing'' is unsuitable to this case, and it is more likely to be the crossing of an active region, with multiple spots simultaneously transited. Given the success of the GP model (see section \ref{sec:trans spec}) we feel it would be unnecessary to attempt to model this explicitly.

\begin{figure*}
\begin{center}
\includegraphics[width=\columnwidth]{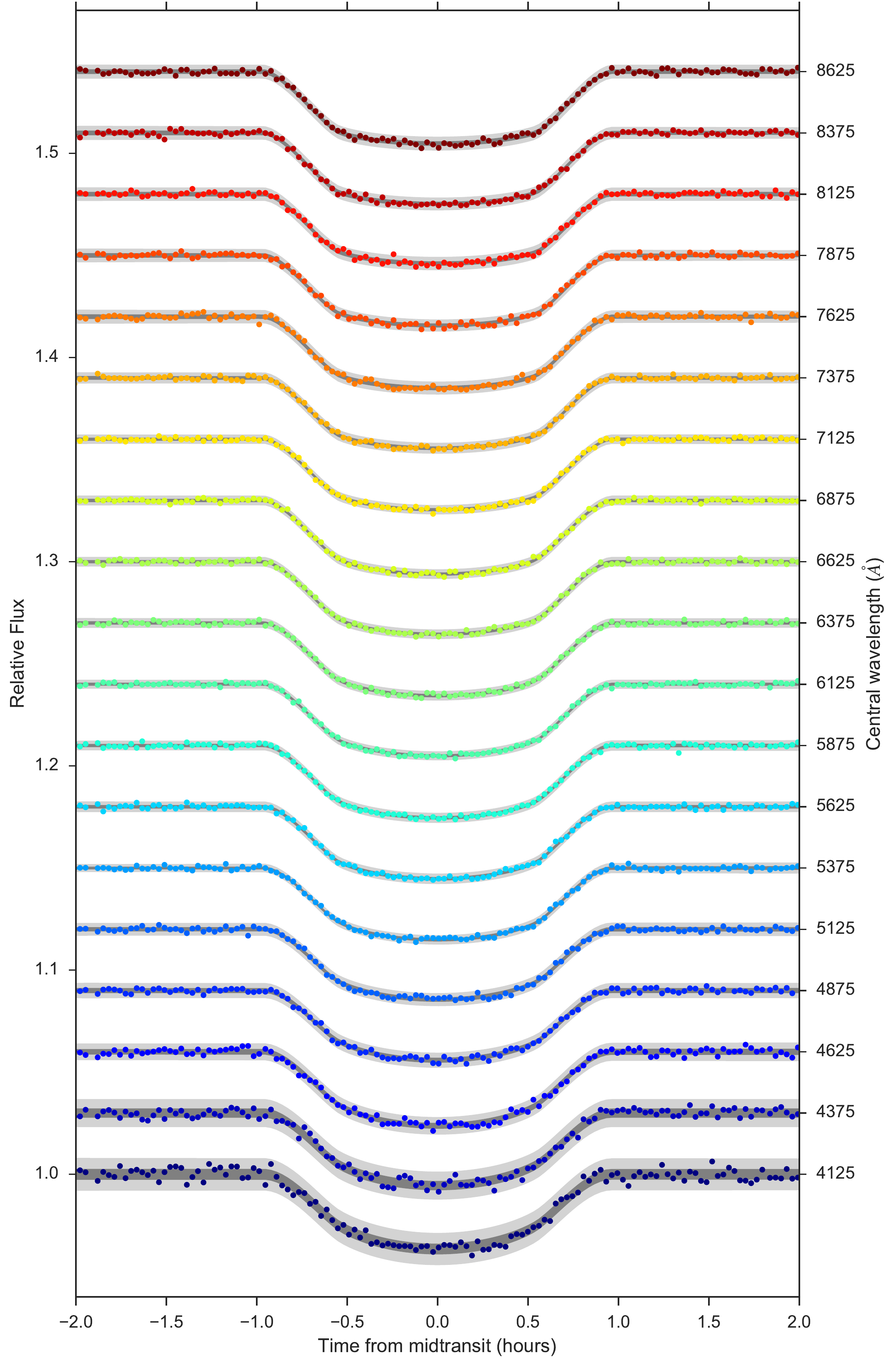}
\includegraphics[width=\columnwidth]{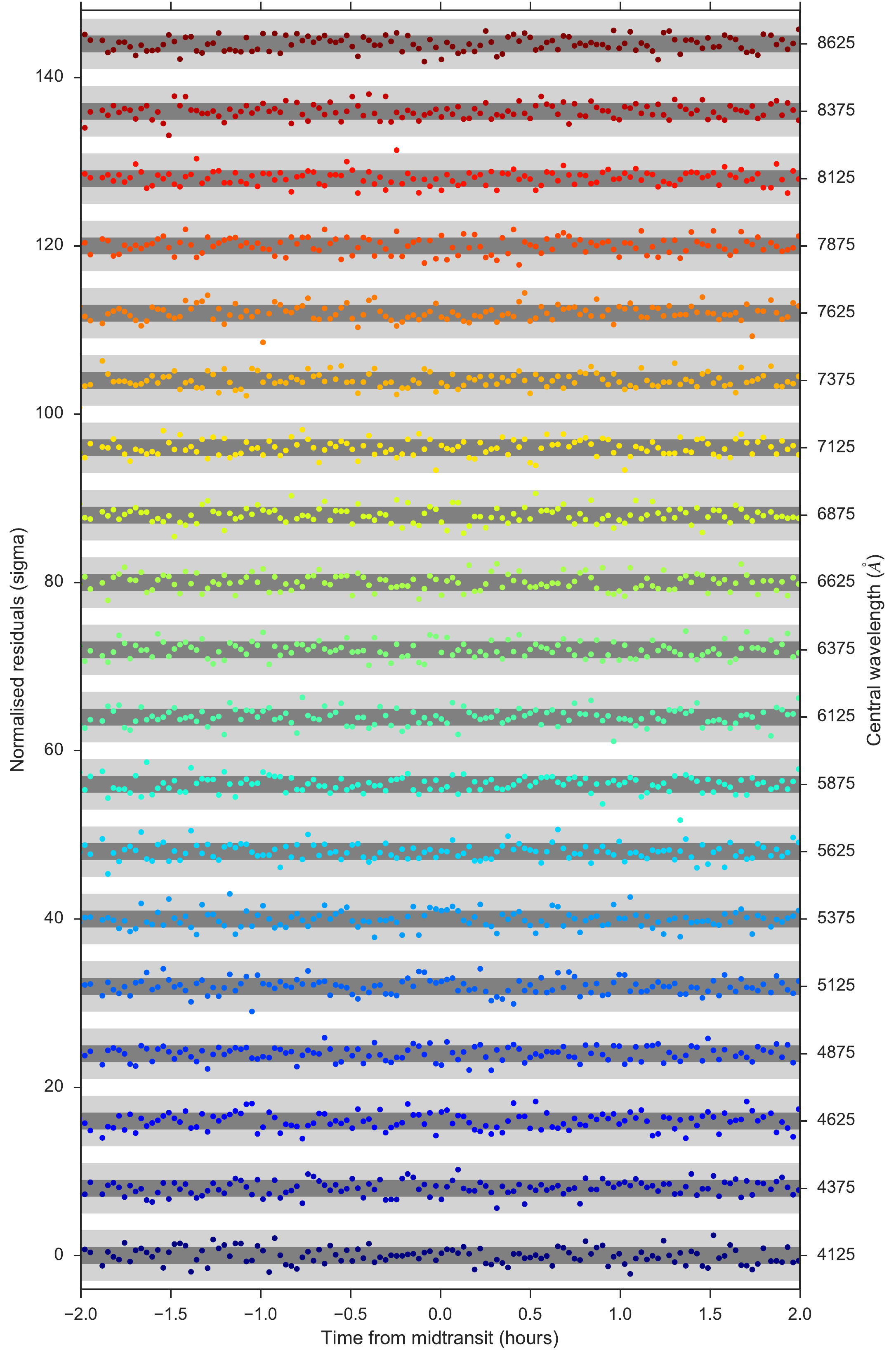}
\caption[Spectral lightcurves for WASP-52 night 1]{\emph{left} spectraly resolved lightcurves and best fitting GP noise models for night 1, after the common mode has been removed. \emph{right} residuals to the best fitting model. Dark grey shaded region indicates $1 \sigma$~ confidence region, dark grey indicates $3 \sigma$~ confidence.}
\label{fig:night1curves}
\end{center}
\end{figure*}

\begin{figure*}
\begin{center}
\includegraphics[width=\columnwidth]{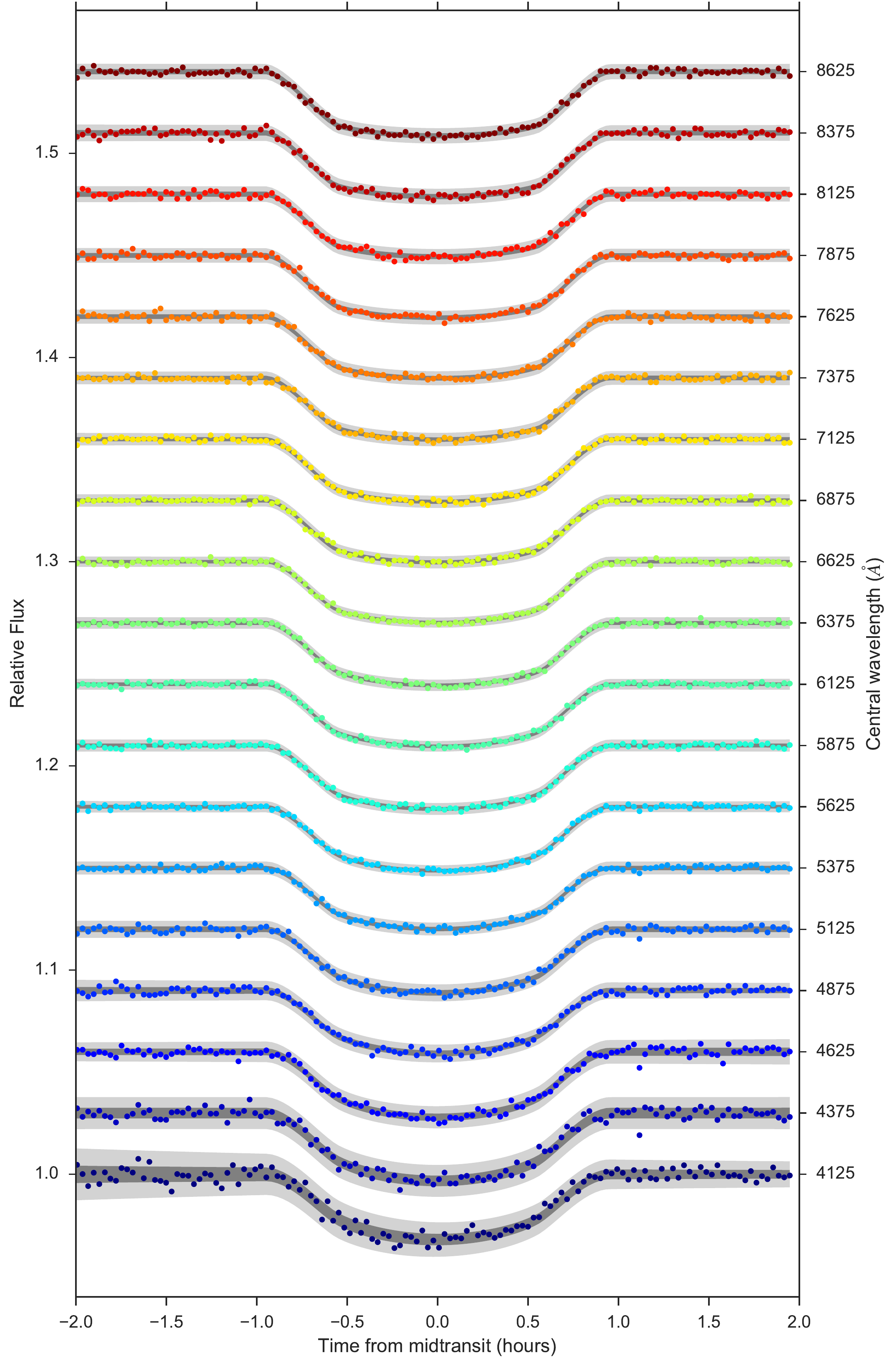}
\includegraphics[width=\columnwidth]{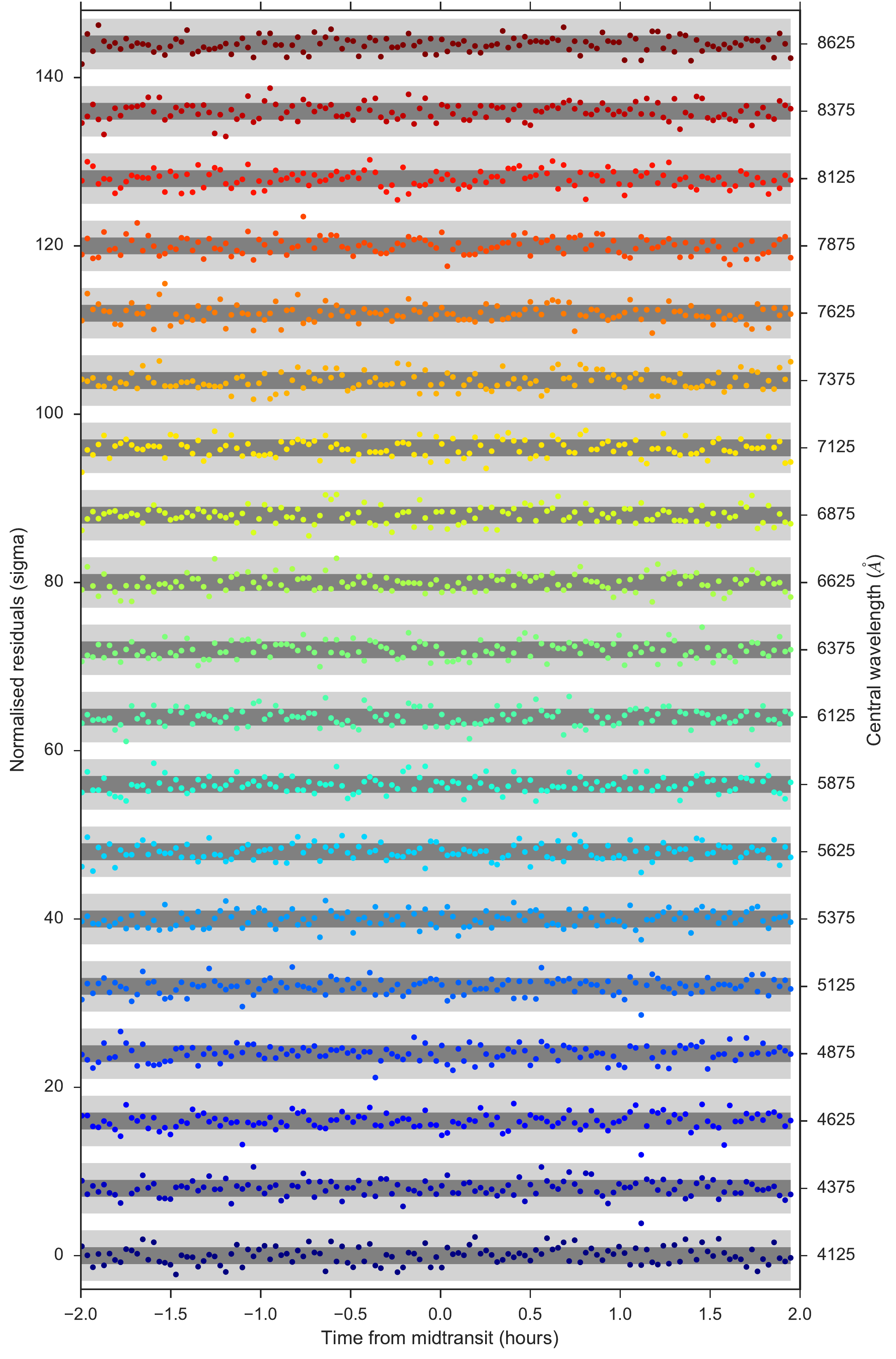}
\caption[Spectral lightcurves for WASP-52 night 2]{\emph{left} spectraly resolved lightcurves and best fitting GP noise models for night 2, after the common mode has been removed. \emph{right} residuals to the best fitting model. Dark grey shaded region indicates $1 \sigma$~ confidence region, dark grey indicates $3 \sigma$~ confidence.}
\label{fig:night2curves}
\end{center}
\end{figure*}

Modelling the spot crossing with the GP is not strictly correct, as the implicit assumption of the Gaussian process approach is that the noise properties are the same for the whole observation window. In principle it would be possible to specify a compound GP that had an additional kernel active only during the transiting portion of the dataset in order to simulate the additional correlated noise component from inhomogeneities of the stellar surface. However, since the event is not visible in the individual spectral bins after common mode removal (see Figures \ref{fig:night1curves} and \ref{fig:night2curves}), we chose not to add additional complexity to the model.

An additional possibility is that this feature is not in fact a spot crossing, but an instrumental systematic with no wavelength dependence, features like this are not uncommon in exoplanet lightcurves. Since the systematic appears to be successfully removed by the common mode noise model, and the system parameters for the two independent nights agree well, the source of the feature is not important, as it will not have an effect on our results.

\section{Results}\label{sec:results}

\subsection{Updated system parameters}\label{sec:system}

\begin{figure*}
\begin{center}
\includegraphics[width=\textwidth]{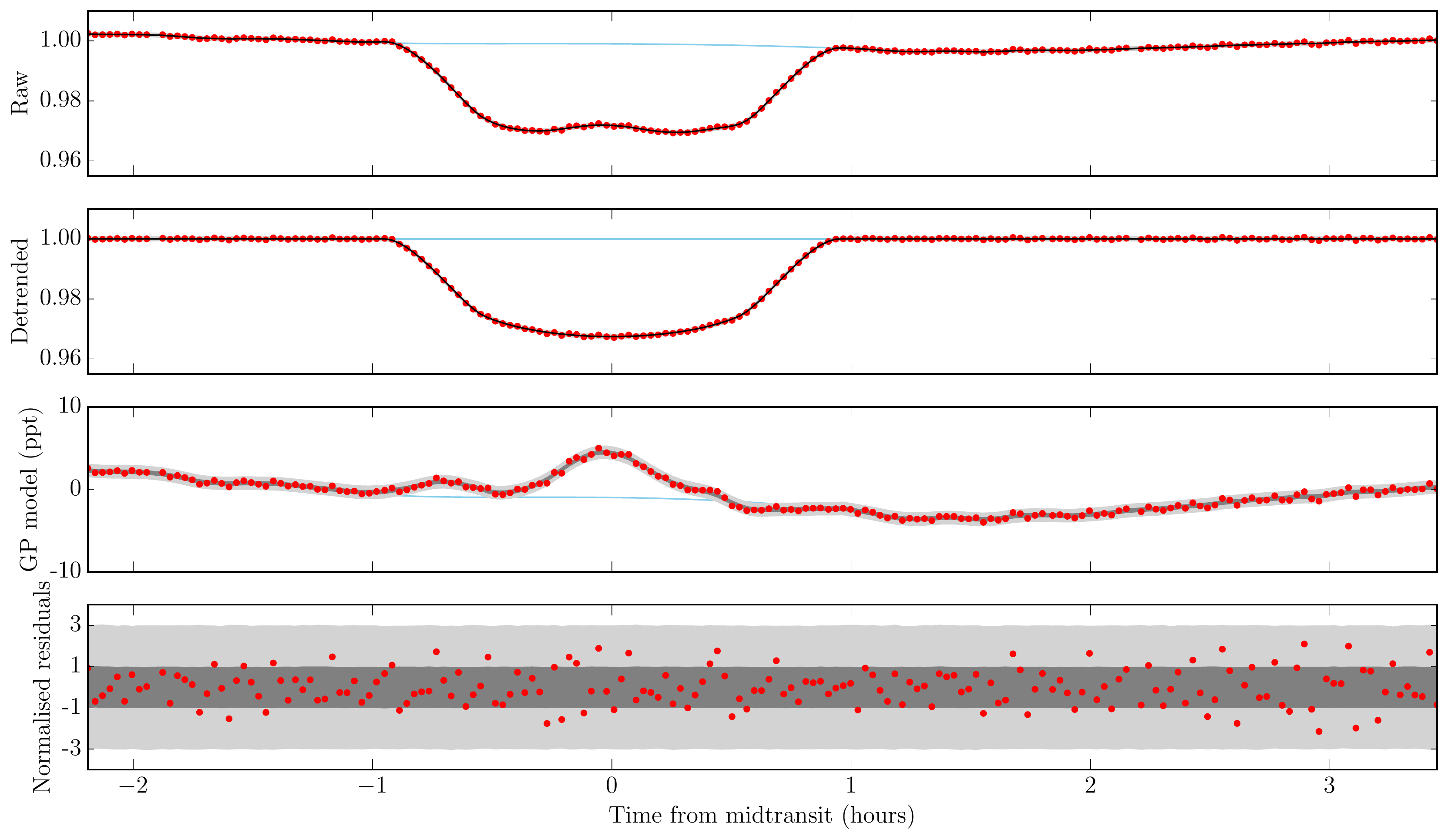}
\includegraphics[width=\textwidth]{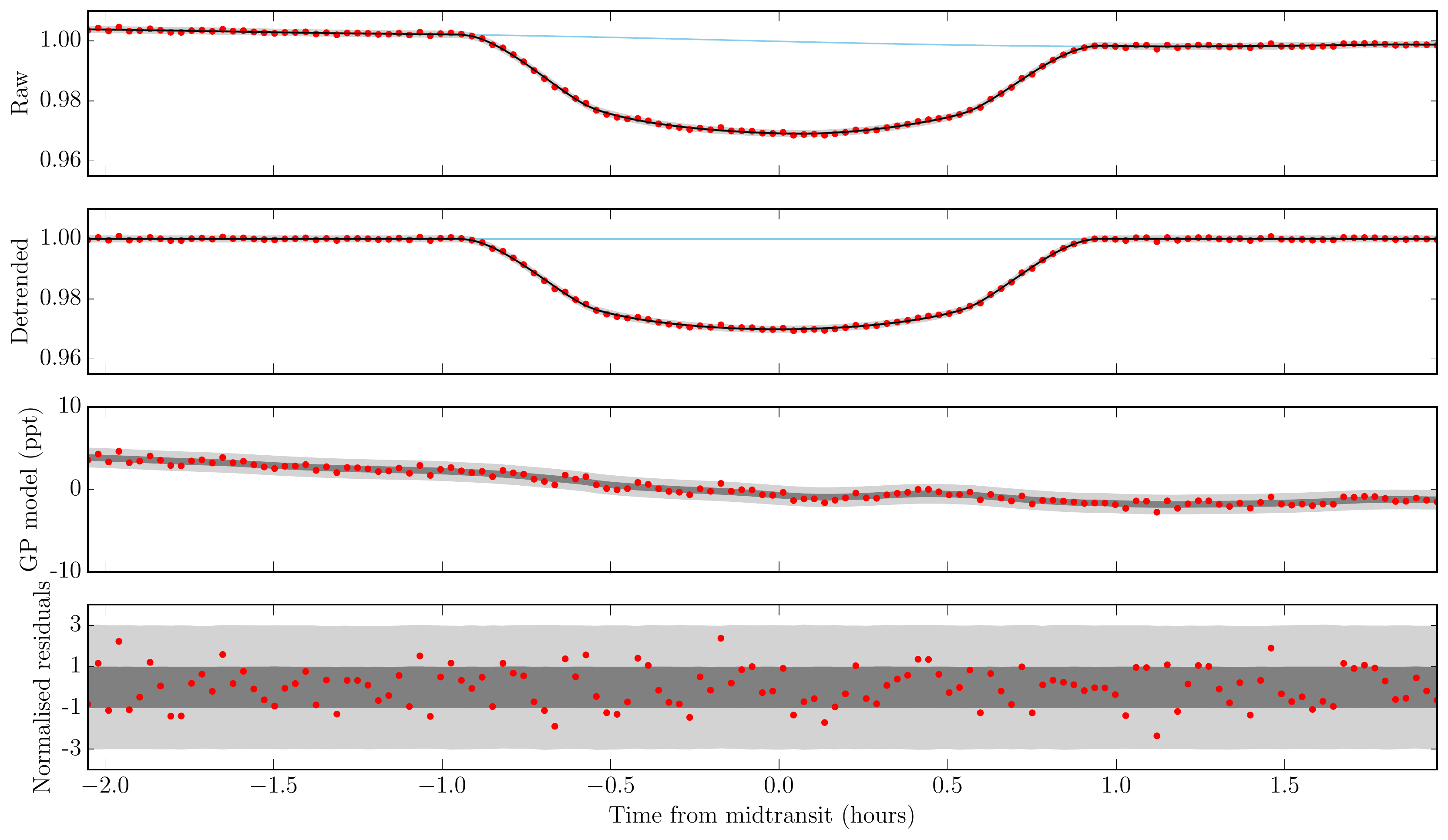}
\caption[White light models for WASP-52b transit]{White light model fits for night 1 (top) and night 2 (bottom). In each, the first panel shows the Gaussian process model fit to the raw data, the second panel shows the model fit to the data with the systematics model removed, the third panel shows the systematic model, and the fourth panel shows the normalised residuals. The dark and light shaded grey regions are the 1 and 3 $\sigma$ confidence regions, respectively.}
\label{fig:whitelcs}
\end{center}
\end{figure*}

We fit the white light curves simultaneously for both nights, using the MCMC GP model, the results are shown in Figure \ref{fig:whitelcs}. We included the transit epoch data from the discovery paper and \citet{Kirk2016} in order to improve the global system parameter estimates. Since the mean stellar surface brightness, and the noise properties are not expected to be the same on both nights, they are allowed to fit separately, but the other system parameters ($a/R_*$, limb darkening coefficients and inclination) are fit simultaneously. Our final system parameters are listed in Table \ref{tab:system}, and agree to within $1\sigma$ with the values in the discovery paper. This provides the system parameters that are fixed during calculation of the spectroscopic lightcurves.

\subsection{Transmission spectra}\label{sec:trans spec}

\begin{figure}
\begin{center}
\includegraphics[width=\columnwidth]{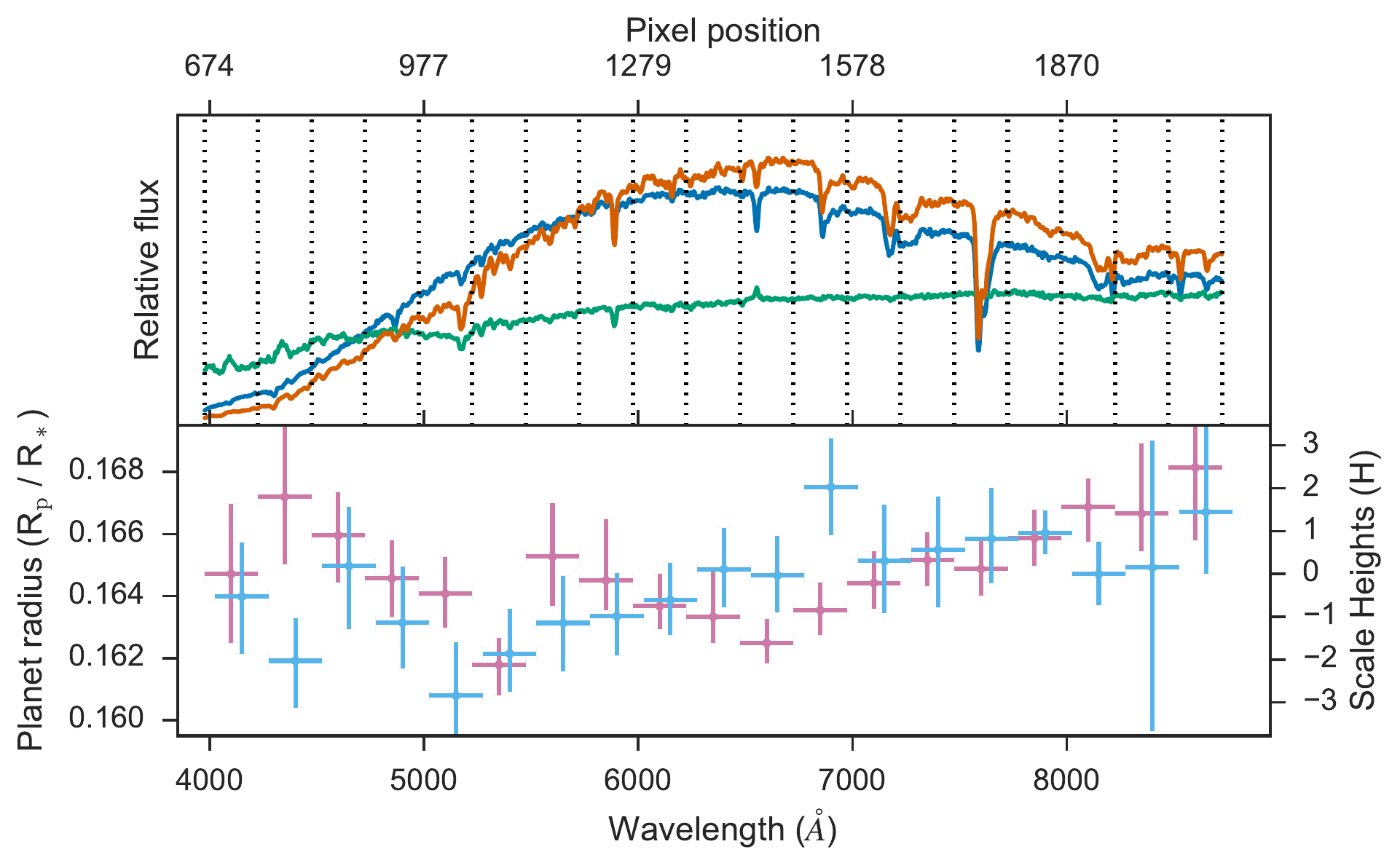}
\caption[Comparison of WASP-52b transmission spectra for two nights]{Top: Spectrum of the target (orange) and comparison (blue) stars and their ratio (green) The 250 \AA\ bins used to generate the spectrum are indicated with dashed lines. Bottom: Transmission spectra from night one (pink) and night two (blue). The absolute offset between the two nights has been removed from the second night, this offset is likely caused by changes is spot coverage. The error bars are the 68\% credible intervals from the MCMC posteriors for each spectral bin. Night 2 has been offset slightly in wavelength for clarity.}
\label{fig:twonights}
\end{center}
\end{figure}

For each 250 \AA\ bin, the median value of the transit depth and the 68\% credible region are calculated from the MCMC posteriors. We present the resulting transmission spectra for the two nights in Table \ref{tab:spectrum} and Figure \ref{fig:twonights}.

\renewcommand{\arraystretch}{1.0}
\begin{table}{
\caption[Transmission spectra for WASP-52b]{Transmission spectrum for WASP-52b. To produce the combined values the absolute offset between Night 1 and Night 2 was removed and a weighted average of the two nights was taken.}
\begin{tabular}{c c c c}
\hline
\hline
Wavelength & & $R_p$/$R_*$ & \\
(\AA) & Night 1 & Night 2 & Combined\\
\hline
4000--4250 & $0.1771\pm0.0018$ & $0.1647\pm0.0022$ & $0.1643\pm0.0014$\\
4250--4500 & $0.1750\pm0.0014$ & $0.1672\pm0.0026$ & $0.1632\pm0.0013$\\
4500--4750 & $0.1780\pm0.0020$ & $0.1660\pm0.0014$ & $0.1656\pm0.0012$\\
4750--5000 & $0.1762\pm0.0016$ & $0.1646\pm0.0012$ & $0.1641\pm0.0010$\\
5000--5250 & $0.1739\pm0.0015$ & $0.1641\pm0.0011$ & $0.1629\pm0.0009$\\
5250--5500 & $0.1752\pm0.0013$ & $0.1618\pm0.0009$ & $0.1619\pm0.0008$\\
5500--5750 & $0.1762\pm0.0015$ & $0.1653\pm0.0017$ & $0.1641\pm0.0011$\\
5750--6000 & $0.1764\pm0.0013$ & $0.1645\pm0.0015$ & $0.1639\pm0.0010$\\
6000--6250 & $0.1769\pm0.0012$ & $0.1637\pm0.0009$ & $0.1638\pm0.0007$\\
6250--6500 & $0.1779\pm0.0013$ & $0.1633\pm0.0012$ & $0.1640\pm0.0009$\\
6500--6750 & $0.1777\pm0.0012$ & $0.1625\pm0.0007$ & $0.1630\pm0.0006$\\
6750--7000 & $0.1806\pm0.0016$ & $0.1635\pm0.0009$ & $0.1645\pm0.0007$\\
7000--7250 & $0.1782\pm0.0017$ & $0.1644\pm0.0009$ & $0.1646\pm0.0008$\\
7250--7500 & $0.1786\pm0.0018$ & $0.1652\pm0.0009$ & $0.1652\pm0.0008$\\
7500--7750 & $0.1789\pm0.0015$ & $0.1649\pm0.0009$ & $0.1651\pm0.0008$\\
7750--8000 & $0.1791\pm0.0007$ & $0.1659\pm0.0009$ & $0.1660\pm0.0006$\\
8000--8250 & $0.1778\pm0.0010$ & $0.1669\pm0.0010$ & $0.1658\pm0.0007$\\
8250--8500 & $0.1780\pm0.0047$ & $0.1667\pm0.0017$ & $0.1664\pm0.0016$\\
8500--8750 & $0.1798\pm0.0024$ & $0.1681\pm0.0022$ & $0.1675\pm0.0016$\\
\end{tabular}
\label{tab:spectrum}
}
\end{table}
\renewcommand{\arraystretch}{1.0}

Having measurements on two nights fitted independently proved valuable, as we were able to use them as an independent test of the assigned error bars. Assuming that the same transmission spectrum is present on both nights, it is expected that the error-normalised deviation of the points from their average will behave like a set of independent Gaussian measurements, with standard deviation proportional to 1/$\sqrt{2}$. Performing this test on the two nights, we find that they do indeed behave as expected for a set of independent measurements of the same distribution, once an absolute depth difference between the two nights has been removed. This indicates that the GP/MCMC method has not underestimated the relative uncertainty in each wavelength bin, and that little common-mode uncertainty remains. It also shows that the spot-crossing event has not introduced any bias to our transmission spectrum.

\begin{figure*}
\begin{center}
\includegraphics[width=\textwidth]{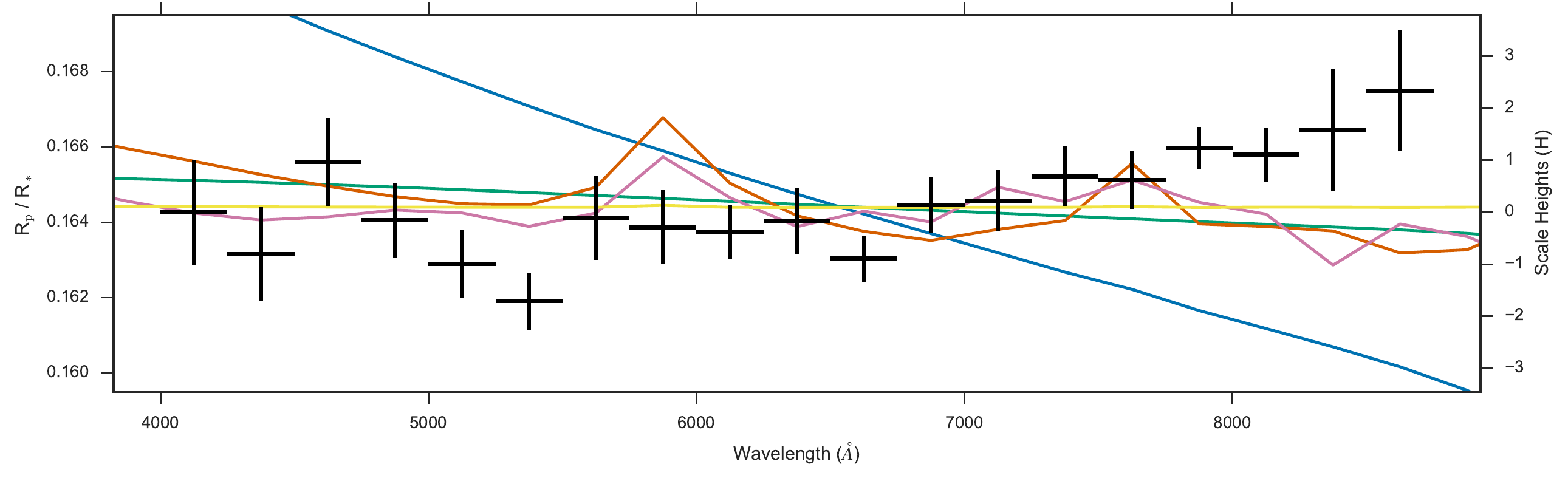}
\caption[Final combined transmission spectrum for WASP-52b]{Final combined transmission spectrum for the two nights. Over plotted are a variety of atmosphere classes. Clear, with solar metalicity and no TiO (orange), a HD\,189733b style Rayleigh scattering slope that covers the entire optical range, with a temperature inversion (blue). A clear atmosphere with a solar abundance of T and V, assuming 100\% is locked up in TiO and VO (pink). A Rayleigh slope with settling (green) representing a hazy atmosphere with larger particles. A grey absorbing cloudbase at 0.1 mbar (yellow).}
\label{fig:spectrum}
\end{center}
\end{figure*}

Since the spectra of the two nights do appear to be describing the same underlying spectrum, and the error bars are reasonable, we use the variance weighted average of the two nights as our final result, shown in Figure \ref{fig:spectrum}

It is worth emphasising that since the common-noise has been removed, these error bars are only correct relative to each other, and are not a valid measurement of uncertainty of the absolute depth.

As a test for sodium absorption, we repeated the process with increasingly narrow down to 50 \AA\ around the 5900 \AA\ doublet, but found no significant difference to the broader bands. Bands narrower than 50 \AA had significantly larger noise, so could not provide further constraints. Whilst the narrow line core of sodium may be present in the spectrum of WASP-52b it is not possible to detect it at the resolution of the instrument. Higher resolution observations are capable of detecting the line core of sodium even above a cloud/haze deck \citep[e.g.][]{Redfield08,Louden2015}.

\section{Discussion}\label{sec:Discussion}

\subsection{Model atmospheres}\label{sec:models}


Inspecting the final transmission spectrum in Figure \ref{fig:spectrum}, the spectrum appears to be largely flat to within a few pressure scale heights, and devoid of any obvious absorption features, though the redmost 4 points are slightly above the median level. Due to the quality and resolution of the available data, a full atmospheric retrieval is not appropriate, especially considering the degeneracies involved in reproducing the flat region of the spectrum (\aprx 4000--7750\AA). Instead we generate a sample of representative atmosphere classes to compare to the data, and qualitatively discuss their applicability to the dataset. A list of all models tested and their goodness of fits are given in Table \ref{tab:atmo_models}.

All of the model atmospheres are generated using the NEMESIS radiative transfer code \citep{Irwin1997,Irwin2008}, with precalculated correlated k-tables described in \citet{Lee2012}. Unless otherwise stated, we assume solar abundances of elements. The only major carbon or oxygen bearing molecule with features in our spectral range is water, for which we assume a volume mixing ratio of $3.83 \times 10^{-4}$, based on the disequilibrium chemistry model of the terminator of HD\,189733b described in \citet{Moses2011}.

We assume simple parametrized Temperature-Pressure profiles generated using the equations presented in \citet{Heng2012}. For all models except the hot haze model, we assume the zero albedo equilibrium temperature for WASP-52b in the case of uniform redistribution, which is 1300 K (using f=0.25 from \citet{Seager2010}). We first calculate a Temperature-Pressure profile appropriate for a clear atmosphere with low levels of short-wave scattering $k_s=6 \times 10^{-4}$ cm$^{-2}$g$^{-1}$ and $\gamma_0=0.06$. 


A clear atmosphere with solar metalicity without TiO and VO absorption is found to be a very poor due to a clear lack of a broadened sodium feature and Rayleigh slope. These features could be masked by high altitude clouds, or de-emphasised by additional molecular absorption.


Several hot Jupiters have been found to have optical spectra that are dominated by Rayleigh slopes, presumably caused by haze. For HD\,189733b, which has a similar equilibrium temperature and parent star spectral type to WASP-52b, \citet{Pont2013} find that the majority of the data taken with \emph{HST/ACS} are well fit by a Rayleigh slope model with an equlibrium temperature of \aprx 1300 K (matching the expected equilibrium temperature of the planet), but bluewards of 6000~\AA\ for data taken with \emph{HST/STIS} the slope is somewhat steeper, having a best fitting temperature of $2100 \pm 500$ K \citep{Sing2011a}. \citet{Pont2013} generate an example analytical Temperature-Pressure profile for HD\,189733b using the equations of \citet{Heng2012} with a strong thermal inversion layer with a thermosphere that reaches 2000 K. This would be capable of reproducing the strong Rayleigh feature seen in the \emph{HST/STIS} data. 
We assume the same parameters as for the T-P profile ($k_s$ = 0.0670, $\gamma_0$ = 10.0, $\eta$ = 0.1) which is equivalent to a geometric albedo of 0.5. For WASP-52b, this results in a thermosphere which reaches 2100 K at 0.1 mbar. The steep Rayleigh slope a haze produces in this hot thermosphere  is clearly a very poor fit to the data, as seen in Figure \ref{fig:spectrum}.

We also test a more subdued Rayleigh slope, using our original non-inverted T-P profile. The slope can be further de-emphasised by assuming slightly larger particles that exhibit settling and hence have a smaller effective scale height in the atmosphere, we parametrise this by using a fractional scale height for the scattering particles that is 0.3 of the gas pressure scale height. This still produces a poor fit that can be confidently ruled out (see Table \ref{tab:atmo_models}).


\citet{Sing2015} show that many hot Jupiters have their optical and near infrared spectral features partially or fully obscured by grey cloud layers, so it seems likely that our spectrum of WASP-52b could also be flattened by high altitude clouds. \citet{Kirk2016} find a flat spectrum compatible with a cloudy interpretation in their analysis of WASP-52b with high precision 3-band photometry. 
To model the effect of clouds we insert opaque layers to the radiative transfer code at 0.1, 1 and 10 mbar. All provide significant improvement over the clear spectrum. The flat spectrum provided by the 0.1 mbar cloud layer is the best fit to the data. We are unable to distinguish between clouds at any higher altitude than 0.1 mbar, as the spectrum is already completely flattened. This is consistent with cloud formation theories for HD\,189733b, which has the same equilibrium temperature as WASP-52b \citep{Lee2015}. These clouds would likely be composed of a range of chemical species, with magnesium silicates as important contributors.


An alternate explanation could be that unresolved molecular features in the planets atmosphere are de-emphasising the Rayleigh slope and sodium features. As an example, we consider an atmosphere with TiO and VO absorption. In hot Jupiter atmospheres T and V in the gas phase are expected to be practically entirely in the oxide forms from chemical equilibrium, but for cooler planets like WASP-52b with equilibrium temperatures below 1600K condensation and cold trapping remove significant fractions from the atmosphere \citep{Fortney2008}. A full treatment of TiO and VO abundance is beyond the scope of this work, instead, we test the effect of several different un-condensed fractions on the spectrum. We assume solar abundances of T and V \citep{Asplund2006}.

A fraction of 1 part per million is indistinguishable from the clear spectrum, but increasing the fraction above this level begins to improve the fit by de-emphasising the Rayleigh slope and sodium feature. We find that an abundance fraction of 0.01 produces a good fit, which is formally indistinguishable from the goodness of fit provided by the cloud dominated model. However, the coolness of the atmosphere of WASP-52b and the lack of firm detections of TiO in other hot Jupiter atmospheres makes this explanation unlikely, so we favour the cloudy model.

A higher resolution search for the presence of the sodium feature (which is still present in the TiO case, but mostly obscured in the cloud-dominated case except for a narrow line core) could potentially help break the degeneracy. A search for the water feature in the infra-red would also be a useful diagnostic, as \citet{Sing2015} shows this is typically obscured in cloudy atmospheres. Our results highlight the need for both low and high resolution observations with a large wavelength coverage to break the degeneracies of exoplanet atmosphere retrieval.


Of the models tested, a flat, cloud dominated spectrum provided the best fit to the full dataset - however the residuals are dominated by the reddest 4 points in the spectrum (7750--9000 \AA). The significant difference in goodness of fit when including these points may be an unexplained systematic error in the data, or indicate the presence of an unknown absorber. 

This feature appears to be present on both nights, showing that it is repeatable, but it may be a common artefact of the instrument or the reduction and analysis.

The feature may have been introduced as an instrumental or atmospheric effect. There are signs of fringing at the reddest ends of the chip It is not on the region of the chip (y pixel $> 2200$), but not in the wavelength region corresponding to the feature. The regions of the chip effected by fringing show a dramatically increased high frequency noise, which is distinct to a persistent increase in radius ratio. The error bars in the 7750--8750 \AA\ region are only somewhat larger than the errors elsewhere in the spectrum. OH Meinel bands are present in the telluric spectrum at these wavelengths, but they are removed by our sky background subtraction and differential photometry. An optical blocking filter was not used when taking these data, so redwards of 6000 \AA\ the spectrum is weakly contaminated by the blue end of the second order spectrum. However, this effect is low order and is not able to increase the observed transit depth above the blue end of the spectrum.

Incorrect limb darkening parameters can cause errors in radius retrieval \citep[e.g.][]{Csizmadia2012a,Espinoza2015} and could introduce a correlated bias to a transmission spectrum. The limb darkening values and errors were generated with Limb Darkening Tool Kit. LDTK fits a parametrised limb darkening profile to PHOENIX-calculated specific intensity spectra \citep{Husser2013}. To test that these model profiles were not influencing the shape of the resulting transmission spectrum the fit was repeated without the LDTK derived priors on the limb darkening parameters. To improve convergence and prevent unphysical profiles we followed the formalism of \cite{Holman2006} and \cite{Burke2008}. We find that without the additional limb darkening constraints the errors on the resulting spectrum are slightly higher, but the shape of the spectrum is preserved. Therefore it is not likely that limb darkening can account for the 4 outlier points.

We can find no obvious systematic reason for this feature in our analysis. While this does not rule additional instrumental effects that we have not considered, it may be a hint of an additional high altitude opacity source. \citet{Nikolov2013a} find a similar redward feature in their \emph{HST/STIS} G750L low-resolution transmission spectrum of HAT-P-1b. A similar effect can also be seen on HD\,209733b \citep{Knutson2007}, WASP-19b with FORS2 \citep{Sedaghati2015} and TrES-3b with OSIRIS \citep{Parviainen2015a}. However, we note that in all these cases, including ours, the unexplained feature is found at the edge of the usable region of the spectra, where the signal to noise is lower and systematics are more likely. This may imply a common instrumental source which we have so far been unable to identify. Alternatively, if this is in fact a real feature, it could be caused by absorption from a molecule not currently included in hot Jupiter atmosphere models. It is also possible that it could be caused by the composition of the particles in the cloud deck, if this layer is vertically extended and has a higher opacity in the red.

\renewcommand{\arraystretch}{1.5}
\begin{center}
\begin{table}{
\caption[Model atmosphere goodness of fits for WASP-52b]{Model atmosphere goodness of fits. The 4000--7750 band has 14 DOF and the 4000--7750 has 18 DOF. The 95\% confidence limit for a $\chi^2_\nu$ distribution with 14 and 18 DOF is 1.69 and 1.60, respectively. No models provide satisfactory fits to the whole range, but both a cloud dominated atmosphere and a clear atmosphere with TiO absorption can provide acceptable fits to the 4000--7750 range.}
\begin{center}
\begin{tabular}{l c c c c c}
\hline
\hline
Model& & $\chi^2$ ($\chi^2_\nu$)&\\
& 4000--7750 \AA\ & &4000--8750 \AA\ \\
\hline
Clear & 29.74 (2.12) & &67.59 (3.76)\\
Hazy (2100 K) &159.92(11.42) & &331.07(18.39)\\
Hazy (1300 K) & 59.9(4.28) & &132.79(7.38)\\
Hazy (with settling) & 26.52(1.89) & &60.43(3.36)\\
TiO and VO ($10^{-6}$) & 29.73 (2.12) & &67.55 (3.75)\\
TiO and VO ($10^{-4}$)& 28.75 (2.05) & &64.19(3.57)\\
TiO and VO ($10^{-2}$)& 16.57 (1.18) & &42.97(2.39)\\
Cloud (10 mbar) & 22.75 (1.62) & &53.45(2.97) \\
Cloud (1 mbar) & 18.82 (1.34) & &41.72(2.32) \\
Cloud (0.1 mbar) & 19.33 (1.38) & &40.89(2.27) \\

\end{tabular}
\end{center}
\label{tab:atmo_models}
}
\end{table}
\end{center}
\renewcommand{\arraystretch}{1.0}

\subsection{Un-occulted spots}\label{unocculted}

WASP-52 is known to be an active star, and as discussed in Section \ref{sec:spotcross}, the star shows periodic modulations in brightness of order \mytilde 1\%, which is likely due to starspots rotating in and out of view. The temperature of the starspots on the very similar K star HD\,189733 were found to be 4250 K by \citet{Sing2011a}. Using this temperature, which is 750 K cooler than the photosphere, gives a spot coverage fraction of about 4\%. This corresponds well to the lower limit of 4\% spot coverage on HD\,189733 calculated by \citet{McCullough2014b}, and is significantly higher than the 0.3\% spot coverage of the Sun at solar maxima \citep{Deming2013}. The value of 4\% spot coverage is an average across an observing season however, and as previously mentioned there was no contemporaneous photometry of WASP-52, so it is possible that the spot coverage was higher than this at the time of these observations.

Un-occulted spots can have an effect on a transmission spectrum, mimicking a Rayleigh scattering slope or molecular features such as water \citep[e.g.][]{McCullough2014b}, so before comparing planetary atmosphere models to the spectra, it is important to quantify what effects un-occulted spots may have.

We simulate the effects of spots on the transmission spectrum using stellar spectra synthesised from a grid of \textsc{atlas9} stellar atmosphere models by \citet{Howarth2011}. We use a log(g) of 4.5 and a temperature of 5000 K for the photosphere, and a range of temperatures for the starspots. The transmission spectrum is generated as
\begin{equation} \label{eq:kmatrix}
\tilde{\delta}(\lambda) = \delta(\lambda)\frac{1}{1 - \eta\left(1 - \frac{F_\lambda(spot)}{F_\lambda(phot)}\right)}
\end{equation}
where $\eta$ is the spotted fraction of the visible stellar surface, $F_\lambda(spot)$ and $F_\lambda(phot)$ are the spectral radiance of the spots and photosphere, $\delta(\lambda)$ is the `true' transit depth at wavelength $\lambda$, here assumed to be a constant value of 0.164, and $\tilde{\delta}(\lambda)$ is the observed value.

\begin{figure}
\begin{center}
\includegraphics[width=\columnwidth]{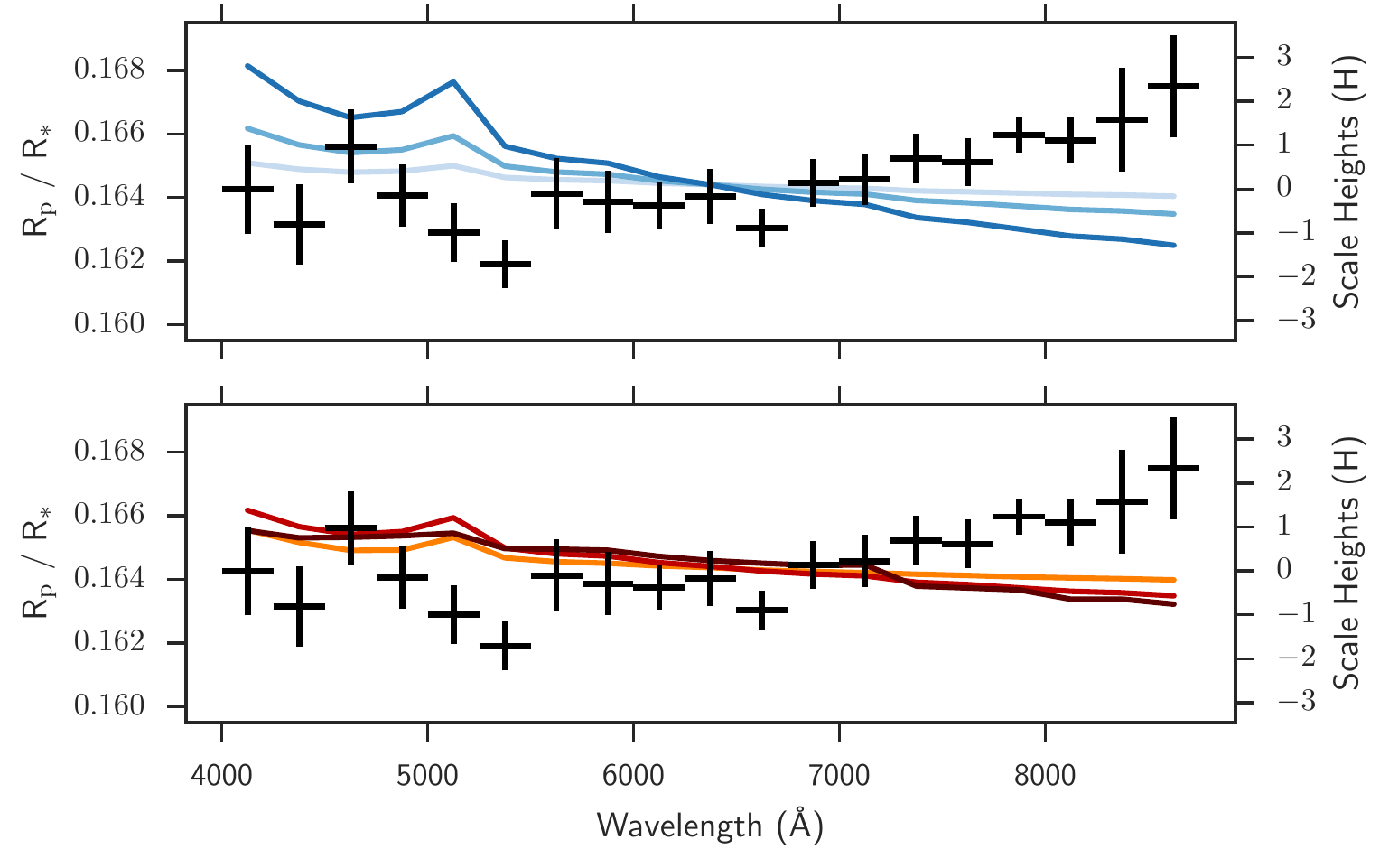}
\caption{The effect of several un-occulted spot scenarios, plotted against the weighted average of the two nights transmission spectrum (black) \emph{Top} Effect of varying the spotty fraction of the star, from light to dark: 0.04,0.1,0.2. all with a spot temperature of 4250 K. \emph{Bottom} Effect of varying the temperature constrast of the starspots, from light to dark: 4750 K, 4250 K, 3750 K. all with a spot coverage fraction of 0.1.}
\label{fig:unocculted}
\end{center}
\end{figure}

The results are plotted in Figure \ref{fig:unocculted} for a range of spot temperatures and coverage fractions. The nominal case of \mytilde 4\% coverage with 4250 K spots does not show any significant features within our level of precision. Increasing the spot contrast and the coverage fraction produces a spectrum with a bluewards slope that does not correspond to the observed spectrum.

We therefore do not expect that un-occulted spots have an appreciable impact on our transmission spectrum.
 

\subsection{Transit timing}\label{sec:ttiming}

\citet{Baluev2015} searched for periodic Transit Timing Variations in 10 exoplanet systems, including WASP-52b and found no evidence of periodic variations. We leverage the high precision of the white noise lightcurves and the longer time baseline between the ephemeris reported in the discovery paper \citep{Hebrard2012} in order to test for secular drifts in the ephemeris of WASP-52b. The error bars on our ephemerides are probably larger than those reported in \citet{Kirk2016} and \citet{Hebrard2012} due to the greater model flexibility allowed by our Gaussian process analysis.

We convert the transit dates to BJD in order to calculate the ephemeris using an online tool\footnote{http://astroutils.astronomy.ohio-state.edu/time/utc2bjd.html} \citep{Eastman2010}.

\citet{Hebrard2012} report their central transit time in HJD, which can introduce errors at the {\raise.17ex\hbox{$\scriptstyle\mathtt{\sim}$}}1 second level compared to calculating BJD$_{TBD}$ directly, so we add this uncertainty in quadrature to their reported errors.

\begin{figure}
\begin{center}
\includegraphics[width=\columnwidth]{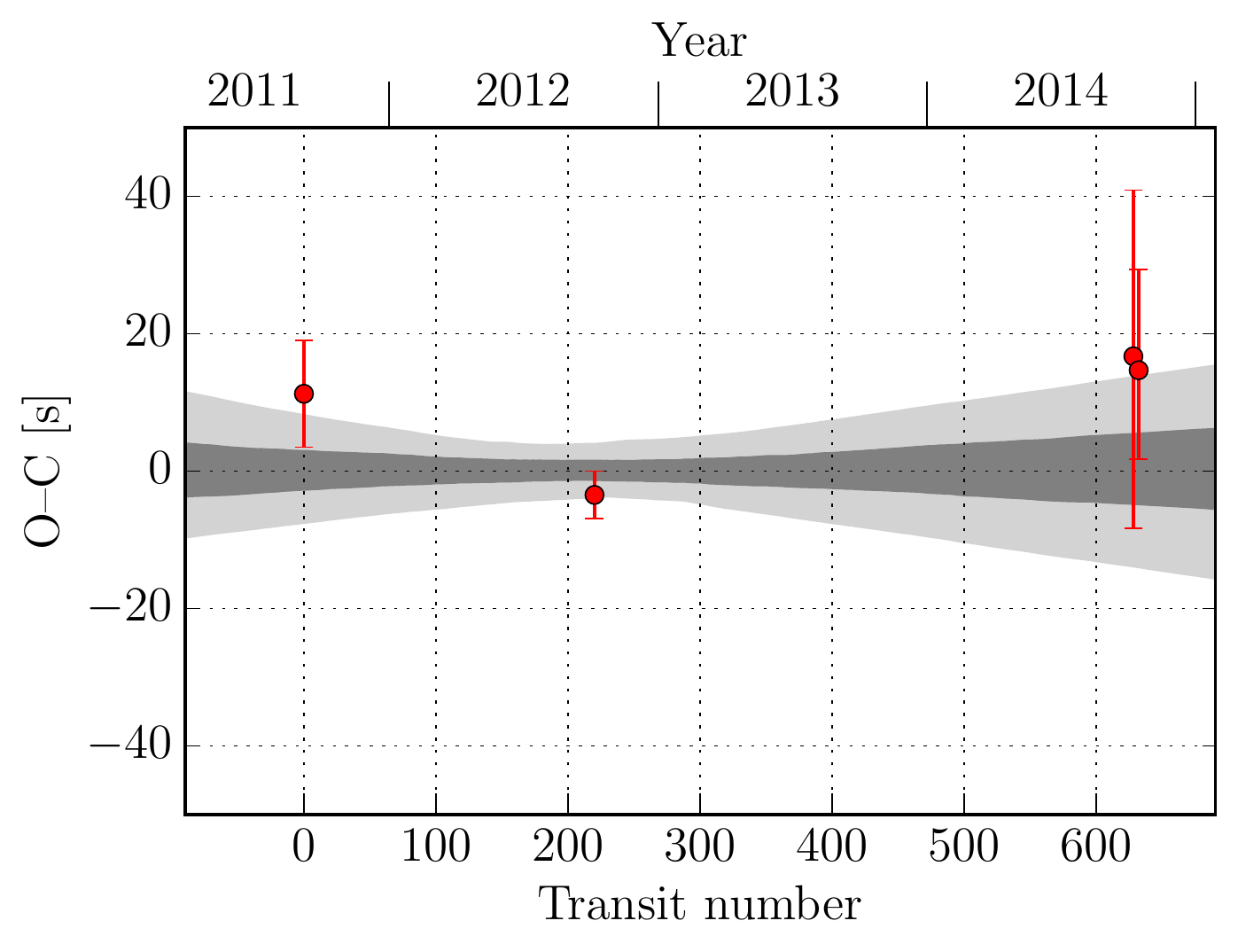}
\caption[Transit timing variations in WASP-52]{Transit timing residuals from a linear ephemeris. The $1\,\sigma$ contour for a linear ephemeris drift model is plotted in grey. No significant deviation from a linear ephemeris is found.}
\label{fig:ttiming}
\end{center}
\end{figure}

We calculated the errors using an MCMC with a chain length of 100,000 steps and a burn in chain of 1000 steps. The best fitting ephemeris is shown in Figure \ref{fig:ttiming}.

A linear ephemeris is formally a poor fit, which could be interpreted as evidence of a non-linear ephemeris in this system. However, we believe it is more likely that previous work may have systematically underestimated the size of their errors by not properly accounting for the presence of correlated noise through Gaussian processes or some other technique. Future measurements of the ephemeris of this system may shed more light on the possibility of additional companions.

The large time baseline (4 years) between the discovery paper and this work allows a significant improvement in the uncertainty of the period. The aforementioned concern over underestimated errors is not particularly important, as the error budget is dominated by the gradient over time periods this long.

\section{Conclusions}

We have performed transmission spectroscopy on WASP-52b, a highly inflated hot Jupiter orbiting an active K star. Our Gaussian process and common noise model approach to modelling resulted in a precise relative transmission spectrum that was consistent across two nights despite the presence of a potential spot. Our errors are on the order of a single scale height.

The possible ``spot-crossing'' feature on the first night appears to be grey, and is not visible in the spectrally resolved light curves after common-noise term removal. This implies that it is either a wavelength independent systematic, or that the temperature contrast of the spot to the rest of the photosphere is high, which is an expected feature for cooler stars such as WASP-52. We have performed a harmonic analysis of the long term photometry of WASP-52, and find that the activity level may be changing over time, with evidence of differential rotation.

We attempted to fit representative model atmospheres to the transmission spectrum, but were unable to find a satisfactory fit to the entire spectral range. For the majority of the spectrum (4000--7750 \AA), a cloud layer at 0.1 mbar provides an acceptable fit to the data, but is inconsistent with a slightly deeper transit at wavelengths $> 7750$ \AA. We explore several different systematic reasons for this excess depth, and are unable to find a compelling cause. If it is a real feature of the atmosphere it may be the result of an additional unknown absorber.

The cloud layer at 0.1 mbar is consistent with cloud formation theories for HD\,189733b, which has the same equilibrium temperature as WASP-52b \citep{Lee2015}. These clouds would likely be composed of a range of chemical species, with magnesium silicates as important contributors.

The transmission spectrum is dramatically different to HD\,189733b, which is dominated by a Rayleigh slope caused by high altitude haze. The difference between these transmission spectra provides further evidence of the rich and as yet unexplained diversity within exoplanet atmospheres.

\section*{Acknowledgements}

We thank the anonymous reviewer, whose comments greatly improved this manuscript. Based on observations made with the WHT operated on the island of La Palma by the Isaac Newton Group in the Spanish Observatorio del Roque de los Muchachos of the Instituto de Astrofisica de Canarias. J.K is supported by an STFC studentship. P.W. and T.L. are supported by STFC consolidated grant (ST/L000733/1).




\bibliographystyle{mnras}
\bibliography{bibliography}

\begin{thebibliography}{}
\makeatletter
\relax
\def\mn@urlcharsother{\let\do\@makeother \do\$\do\&\do\#\do\^\do\_\do\%\do\~}
\def\mn@doi{\begingroup\mn@urlcharsother \@ifnextchar [ {\mn@doi@}
  {\mn@doi@[]}}
\def\mn@doi@[#1]#2{\def\@tempa{#1}\ifx\@tempa\@empty \href
  {http://dx.doi.org/#2} {doi:#2}\else \href {http://dx.doi.org/#2} {#1}\fi
  \endgroup}
\def\mn@eprint#1#2{\mn@eprint@#1:#2::\@nil}
\def\mn@eprint@arXiv#1{\href {http://arxiv.org/abs/#1} {{\tt arXiv:#1}}}
\def\mn@eprint@dblp#1{\href {http://dblp.uni-trier.de/rec/bibtex/#1.xml}
  {dblp:#1}}
\def\mn@eprint@#1:#2:#3:#4\@nil{\def\@tempa {#1}\def\@tempb {#2}\def\@tempc
  {#3}\ifx \@tempc \@empty \let \@tempc \@tempb \let \@tempb \@tempa \fi \ifx
  \@tempb \@empty \def\@tempb {arXiv}\fi \@ifundefined
  {mn@eprint@\@tempb}{\@tempb:\@tempc}{\expandafter \expandafter \csname
  mn@eprint@\@tempb\endcsname \expandafter{\@tempc}}}

\bibitem[\protect\citeauthoryear{Ambikasaran, Foreman-Mackey, Greengard, Hogg
  \& O'Neil}{Ambikasaran et~al.}{2016}]{Ambikasaran2016}
Ambikasaran S.,  Foreman-Mackey D.,  Greengard L.,  Hogg D.~W.,   O'Neil M.,
  2016, \mn@doi [IEEE Trans. Pattern Anal. Mach. Intell.]
  {10.1109/TPAMI.2015.2448083}, 38, 252

\bibitem[\protect\citeauthoryear{Asplund, Grevesse  \& {Jacques
  Sauval}}{Asplund et~al.}{2006}]{Asplund2006}
Asplund M.,  Grevesse N.,   {Jacques Sauval} A.,  2006, \mn@doi [Nucl. Phys. A]
  {10.1016/j.nuclphysa.2005.06.010}, 777, 1

\bibitem[\protect\citeauthoryear{Baluev et~al.,}{Baluev
  et~al.}{2015}]{Baluev2015}
Baluev R.~V.,  et~al., 2015, \mn@doi [MNRAS] {10.1093/mnras/stv788}, 450, 3101

\bibitem[\protect\citeauthoryear{Bento et~al.,}{Bento et~al.}{2014}]{Bento2013}
Bento J.,  et~al., 2014, \mn@doi [MNRAS] {10.1093/mnras/stt1979}, 437, 1511

\bibitem[\protect\citeauthoryear{Borsa, Rainer  \& Poretti}{Borsa
  et~al.}{2016}]{Borsa2016}
Borsa F.,  Rainer M.,   Poretti E.,  2016, \mn@doi [A{\&}A]
  {10.1051/0004-6361/201628334}, 590, A84

\bibitem[\protect\citeauthoryear{Burke et~al.,}{Burke et~al.}{2008}]{Burke2008}
Burke C.~J.,  et~al., 2008, \mn@doi [ApJ] {10.1086/591497}, 686, 1331

\bibitem[\protect\citeauthoryear{{Collier Cameron} et~al.,}{{Collier Cameron}
  et~al.}{2009}]{CollierCameron2009}
{Collier Cameron} A.,  et~al., 2009, \mn@doi [MNRAS]
  {10.1111/j.1365-2966.2009.15476.x}, 400, 451

\bibitem[\protect\citeauthoryear{Csizmadia, Pasternacki, Dreyer, Cabrera,
  Erikson  \& Rauer}{Csizmadia et~al.}{2012}]{Csizmadia2012a}
Csizmadia S.,  Pasternacki T.,  Dreyer C.,  Cabrera J.,  Erikson a.,   Rauer
  H.,  2012, \mn@doi [A{\&}A] {10.1051/0004-6361/201219888}, 549, A9

\bibitem[\protect\citeauthoryear{Deming et~al.,}{Deming
  et~al.}{2013}]{Deming2013}
Deming D.,  et~al., 2013, \mn@doi [ApJ] {10.1088/0004-637X/774/2/95}, 774, 95

\bibitem[\protect\citeauthoryear{Eastman, Siverd  \& Gaudi}{Eastman
  et~al.}{2010}]{Eastman2010}
Eastman J.,  Siverd R.,   Gaudi B.~S.,  2010, \mn@doi [PASP] {10.1086/655938},
  122, 935

\bibitem[\protect\citeauthoryear{Espinoza \& Jordan}{Espinoza \&
  Jordan}{2015}]{Espinoza2015}
Espinoza N.,  Jordan A.,  2015, \mn@doi [Mnras] {10.1093/mnras/stv744}, 450,
  1879

\bibitem[\protect\citeauthoryear{Evans, Aigrain, Gibson, Barstow, Amundsen,
  Tremblin  \& Mourier}{Evans et~al.}{2015}]{Evans2015}
Evans T.~M.,  Aigrain S.,  Gibson N.,  Barstow J.~K.,  Amundsen D.~S.,
  Tremblin P.,   Mourier P.,  2015, \mn@doi [MNRAS] {10.1093/mnras/stv910},
  451, 680

\bibitem[\protect\citeauthoryear{Evans et~al.,}{Evans et~al.}{2016}]{Evans2016}
Evans T.~M.,  et~al., 2016, \mn@doi [ApJ] {10.3847/2041-8205/822/1/L4}, 822, L4

\bibitem[\protect\citeauthoryear{Faedi et~al.,}{Faedi et~al.}{2011}]{Faedi2011}
Faedi F.,  et~al., 2011, \mn@doi [EPJ Web Conf.] {10.1051/epjconf/20101101003},
  11, 01003

\bibitem[\protect\citeauthoryear{Foreman-Mackey, Hogg, Lang  \&
  Goodman}{Foreman-Mackey et~al.}{2013}]{Foreman-Mackey2013}
Foreman-Mackey D.,  Hogg D.~W.,  Lang D.,   Goodman J.,  2013, \mn@doi [PASP]
  {10.1086/670067}, 125, 306

\bibitem[\protect\citeauthoryear{Fortney, Lodders, Marley  \& Freedman}{Fortney
  et~al.}{2008}]{Fortney2008}
Fortney J.~J.,  Lodders K.,  Marley M.~S.,   Freedman R.~S.,  2008, \mn@doi
  [ApJ] {10.1086/528370}, 678, 1419

\bibitem[\protect\citeauthoryear{Gibson, Aigrain, Barstow, Evans, Fletcher  \&
  Irwin}{Gibson et~al.}{2013a}]{Gibson2013a}
Gibson N.~P.,  Aigrain S.,  Barstow J.~K.,  Evans T.~M.,  Fletcher L.~N.,
  Irwin P. G.~J.,  2013a, \mn@doi [MNRAS] {10.1093/mnras/sts307}, 428, 3680

\bibitem[\protect\citeauthoryear{Gibson, Aigrain, Barstow, Evans, Fletcher  \&
  Irwin}{Gibson et~al.}{2013b}]{Gibson2013}
Gibson N.~P.,  Aigrain S.,  Barstow J.~K.,  Evans T.~M.,  Fletcher L.~N.,
  Irwin P. G.~J.,  2013b, \mn@doi [MNRAS] {10.1093/mnras/stt1783}, 436, 2974

\bibitem[\protect\citeauthoryear{Gloria, Snellen  \& Albrecht}{Gloria
  et~al.}{2015}]{Gloria2015}
Gloria E.~D.,  Snellen I. A.~G.,   Albrecht S.,  2015, 84, 1

\bibitem[\protect\citeauthoryear{Goodman \& Weare}{Goodman \&
  Weare}{2010}]{Goodman2010}
Goodman J.,  Weare J.,  2010, \mn@doi [Commun. Appl. Math. Comput. Sci.]
  {10.2140/camcos.2010.5.65}, 5, 65

\bibitem[\protect\citeauthoryear{Hebrard et~al.,}{Hebrard
  et~al.}{2012}]{Hebrard2012}
Hebrard G.,  et~al., 2012, \mn@doi [A{\&}A] {10.1051/0004-6361/201220363}, 134,
  13

\bibitem[\protect\citeauthoryear{Heng, Hayek, Pont  \& Sing}{Heng
  et~al.}{2012}]{Heng2012}
Heng K.,  Hayek W.,  Pont F.,   Sing D.~K.,  2012, \mn@doi [MNRAS]
  {10.1111/j.1365-2966.2011.19943.x}, 420, 20

\bibitem[\protect\citeauthoryear{Holman et~al.,}{Holman
  et~al.}{2006}]{Holman2006}
Holman M.~J.,  et~al., 2006, \mn@doi [ApJ] {10.1086/508155}, 652, 1715

\bibitem[\protect\citeauthoryear{Howarth}{Howarth}{2011}]{Howarth2011}
Howarth I.~D.,  2011, \mn@doi [MNRAS] {10.1111/j.1365-2966.2011.18122.x}, 413,
  1515

\bibitem[\protect\citeauthoryear{Husser, {Wende-von Berg}, Dreizler, Homeier,
  Reiners, Barman  \& Hauschildt}{Husser et~al.}{2013}]{Husser2013}
Husser T.-O.,  {Wende-von Berg} S.,  Dreizler S.,  Homeier D.,  Reiners a.,
  Barman T.,   Hauschildt P.~H.,  2013, \mn@doi [A{\&}A]
  {10.1051/0004-6361/201219058}, 553, A6

\bibitem[\protect\citeauthoryear{Irwin, Calcutt  \& Taylor}{Irwin
  et~al.}{1997}]{Irwin1997}
Irwin P.,  Calcutt S.,   Taylor F.,  1997, \mn@doi [Adv. Sp. Res.]
  {10.1016/S0273-1177(97)00266-4}, 19, 1149

\bibitem[\protect\citeauthoryear{Irwin et~al.,}{Irwin et~al.}{2008}]{Irwin2008}
Irwin P.,  et~al., 2008, \mn@doi [J. Quant. Spectrosc. Radiat. Transf.]
  {10.1016/j.jqsrt.2007.11.006}, 109, 1136

\bibitem[\protect\citeauthoryear{Kirk, Wheatley, Louden, Littlefair,
  Copperwheat, Armstrong, Marsh  \& Dhillon}{Kirk et~al.}{2016}]{Kirk2016}
Kirk J.,  Wheatley P.~J.,  Louden T.,  Littlefair S.~P.,  Copperwheat C.~M.,
  Armstrong D.~J.,  Marsh T.~R.,   Dhillon V.~S.,  2016, \mn@doi [MNRAS]
  {10.1093/mnras/stw2205}, 463, 2922

\bibitem[\protect\citeauthoryear{Knutson, Charbonneau, Noyes, Brown  \&
  Gilliland}{Knutson et~al.}{2007}]{Knutson2007}
Knutson H.~A.,  Charbonneau D.,  Noyes R.~W.,  Brown T.~M.,   Gilliland R.~L.,
  2007, \mn@doi [ApJ] {10.1086/510111}, 655, 564

\bibitem[\protect\citeauthoryear{Knutson, Charbonneau, Allen, Burrows  \&
  Megeath}{Knutson et~al.}{2008}]{Knutson2008a}
Knutson H.~A.,  Charbonneau D.,  Allen L.,  Burrows A.,   Megeath S.~T.,  2008,
  \mn@doi [ApJ] {10.1086/523894}, 673, 526

\bibitem[\protect\citeauthoryear{Kreidberg et~al.,}{Kreidberg
  et~al.}{2014}]{Kreidberg2014a}
Kreidberg L.,  et~al., 2014, \mn@doi [Nature] {10.1038/nature12888}, 505, 69

\bibitem[\protect\citeauthoryear{Lee \& Juncher}{Lee \&
  Juncher}{2015}]{Lee2015}
Lee G.,  Juncher D.,  2015, ] {10.1051/0004-6361/201525982}, 12, 1

\bibitem[\protect\citeauthoryear{Lee, Fletcher  \& Irwin}{Lee
  et~al.}{2012}]{Lee2012}
Lee J.-M.,  Fletcher L.~N.,   Irwin P. G.~J.,  2012, \mn@doi [MNRAS]
  {10.1111/j.1365-2966.2011.20013.x}, 420, 170

\bibitem[\protect\citeauthoryear{Lendl et~al.,}{Lendl et~al.}{2016}]{Lendl2015}
Lendl M.,  et~al., 2016, \mn@doi [A{\&}A] {10.1051/0004-6361/201527594}, 587,
  A67

\bibitem[\protect\citeauthoryear{Line \& Yung}{Line \& Yung}{2013}]{Line2013}
Line M.~R.,  Yung Y.~L.,  2013, \mn@doi [ApJ] {10.1088/0004-637X/779/1/3}, 779,
  3

\bibitem[\protect\citeauthoryear{Louden \& Wheatley}{Louden \&
  Wheatley}{2015}]{Louden2015}
Louden T.,  Wheatley P.~J.,  2015, \mn@doi [ApJ] {10.1088/2041-8205/814/2/L24},
  814, L24

\bibitem[\protect\citeauthoryear{Mandel \& Agol}{Mandel \&
  Agol}{2002}]{Mandel2002}
Mandel K.,  Agol E.,  2002, \mn@doi [ApJ] {10.1086/345520}, 580, L171

\bibitem[\protect\citeauthoryear{Maxted et~al.,}{Maxted
  et~al.}{2011}]{Maxted2011}
Maxted P. F.~L.,  et~al., 2011, \mn@doi [PASP] {10.1086/660007}, 123, 547

\bibitem[\protect\citeauthoryear{McCullough, Crouzet, Deming  \&
  Madhusudhan}{McCullough et~al.}{2014}]{McCullough2014b}
McCullough P.~R.,  Crouzet N.,  Deming D.,   Madhusudhan N.,  2014, \mn@doi
  [ApJ] {10.1088/0004-637X/791/1/55}, 791, 55

\bibitem[\protect\citeauthoryear{Moehler, Freudling, M{\o}ller, Patat,
  Rupprecht  \& O'Brien}{Moehler et~al.}{2010}]{Moehler2010}
Moehler S.,  Freudling W.,  M{\o}ller P.,  Patat F.,  Rupprecht G.,   O'Brien
  K.,  2010, \mn@doi [Pasp] {10.1086/649963}, 122, 93

\bibitem[\protect\citeauthoryear{Moses et~al.,}{Moses et~al.}{2011}]{Moses2011}
Moses J.~I.,  et~al., 2011, \mn@doi [ApJ] {10.1088/0004-637X/737/1/15}, 737, 15

\bibitem[\protect\citeauthoryear{Nikolov et~al.,}{Nikolov
  et~al.}{2014}]{Nikolov2013a}
Nikolov N.,  et~al., 2014, \mn@doi [MNRAS] {10.1093/mnras/stt1859}, 437, 46

\bibitem[\protect\citeauthoryear{Parmentier, Showman  \& Lian}{Parmentier
  et~al.}{2013}]{Parmentier2013}
Parmentier V.,  Showman A.~P.,   Lian Y.,  2013, \mn@doi [A{\&}A]
  {10.1051/0004-6361/201321132}, 558, A91

\bibitem[\protect\citeauthoryear{Parmentier, Fortney, Showman, Morley  \&
  Marley}{Parmentier et~al.}{2016}]{Parmentier2016}
Parmentier V.,  Fortney J.~J.,  Showman A.~P.,  Morley C.~V.,   Marley M.~S.,
  2016

\bibitem[\protect\citeauthoryear{Parviainen \& Aigrain}{Parviainen \&
  Aigrain}{2015}]{Parviainen2015}
Parviainen H.,  Aigrain S.,  2015, \mn@doi [MNRAS] {10.1093/mnras/stv1857},
  453, 3822

\bibitem[\protect\citeauthoryear{Parviainen, Pall{\'{e}}, Nortmann, Nowak, Iro,
  Murgas  \& Aigrain}{Parviainen et~al.}{2016}]{Parviainen2015a}
Parviainen H.,  Pall{\'{e}} E.,  Nortmann L.,  Nowak G.,  Iro N.,  Murgas F.,
  Aigrain S.,  2016, \mn@doi [A{\&}A] {10.1051/0004-6361/201526313}, 585, A114

\bibitem[\protect\citeauthoryear{Pont, Sing, Gibson, Aigrain, Henry  \&
  Husnoo}{Pont et~al.}{2013}]{Pont2013}
Pont F.,  Sing D.~K.,  Gibson N.~P.,  Aigrain S.,  Henry G.,   Husnoo N.,
  2013, \mn@doi [MNRAS] {10.1093/mnras/stt651}, 432, 2917

\bibitem[\protect\citeauthoryear{Redfield, Endl, Cochran  \&
  Koesterke}{Redfield et~al.}{2008}]{Redfield08}
Redfield S.,  Endl M.,  Cochran W.~D.,   Koesterke L.,  2008, \mn@doi [ApJ]
  {10.1086/527475}, 673, L87

\bibitem[\protect\citeauthoryear{Seager}{Seager}{2010}]{Seager2010}
Seager S.,  2010, {Exoplanet Atmospheres: Physical Processes}.
Princeton University Press, NJ

\bibitem[\protect\citeauthoryear{Sedaghati, Boffin, Csizmadia, Gibson, Kabath,
  Mallonn  \& {Van den Ancker}}{Sedaghati et~al.}{2015}]{Sedaghati2015}
Sedaghati E.,  Boffin H. M.~J.,  Csizmadia S.,  Gibson N.,  Kabath P.,  Mallonn
  M.,   {Van den Ancker} M.~E.,  2015, \mn@doi [A{\&}A]
  {10.1051/0004-6361/201525822}, 576, L11

\bibitem[\protect\citeauthoryear{Sing, Vidal‐Madjar, D{\'{e}}sert,
  {Lecavelier des Etangs}  \& Ballester}{Sing et~al.}{2008}]{Sing2008a}
Sing D.~K.,  Vidal‐Madjar A.,  D{\'{e}}sert J.,  {Lecavelier des Etangs} A.,
   Ballester G.,  2008, \mn@doi [ApJ] {10.1086/590075}, 686, 658

\bibitem[\protect\citeauthoryear{Sing et~al.,}{Sing et~al.}{2011}]{Sing2011a}
Sing D.~K.,  et~al., 2011, \mn@doi [MNRAS] {10.1111/j.1365-2966.2011.19142.x},
  416, 1443

\bibitem[\protect\citeauthoryear{Sing et~al.,}{Sing et~al.}{2012}]{Sing2012}
Sing D.~K.,  et~al., 2012, \mn@doi [MNRAS] {10.1111/j.1365-2966.2012.21938.x},
  426, 1663

\bibitem[\protect\citeauthoryear{Sing et~al.,}{Sing et~al.}{2014}]{Sing2014}
Sing D.~K.,  et~al., 2014, \mn@doi [MNRAS] {10.1093/mnras/stu2279}, 446, 2428

\bibitem[\protect\citeauthoryear{Sing et~al.,}{Sing et~al.}{2016}]{Sing2015}
Sing D.~K.,  et~al., 2016, \mn@doi [Nature] {10.1038/nature16068}, 529, 59

\bibitem[\protect\citeauthoryear{Vidal-Madjar, {Des Etangs}, D{\'{e}}sert,
  Ballester, Ferlet, H{\'{e}}brard  \& Mayor}{Vidal-Madjar
  et~al.}{2003}]{Vidal-Madjar2003}
Vidal-Madjar A.,  {Des Etangs} A.~L.,  D{\'{e}}sert J.-M.,  Ballester G.~E.,
  Ferlet R.,  H{\'{e}}brard G.,   Mayor M.,  2003, \mn@doi [Nature]
  {10.1038/nature01448}, 422, 143

\bibitem[\protect\citeauthoryear{Wakeford \& Sing}{Wakeford \&
  Sing}{2015}]{Wakeford2015}
Wakeford H.~R.,  Sing D.~K.,  2015, \mn@doi [A{\&}A]
  {10.1051/0004-6361/201424207}, 573, A122

\bibitem[\protect\citeauthoryear{Wakeford, Sing, Evans, Deming  \&
  Mandell}{Wakeford et~al.}{2016}]{Wakeford2016}
Wakeford H.~R.,  Sing D.~K.,  Evans T.,  Deming D.,   Mandell A.,  2016,
  \mn@doi [ApJ] {10.3847/0004-637X/819/1/10}, 819, 10

\bibitem[\protect\citeauthoryear{Zechmeister \& K{\"{u}}rster}{Zechmeister \&
  K{\"{u}}rster}{2009}]{Zechmeister2009}
Zechmeister M.,  K{\"{u}}rster M.,  2009, \mn@doi [eprint arXiv]
  {10.1051/0004-6361:200811296}, 0901, 2573

\makeatother
\end{thebibliography}








\bsp	
\label{lastpage}
\end{document}